\documentclass{emulateapj}

\newcommand{\C}{\mbox{C\,\raisebox{+.38ex}{$_{\rm{IV}}$}}}
\newcommand{\Ox}{\mbox{O\,\raisebox{+.38ex}{$_{\rm{III}}$}}}
\newcommand{\Fe}{\mbox{Fe\,\raisebox{+.38ex}{$_{\rm{II}}$}}}
\newcommand{\MBH}{M_{BH}}

\shorttitle{The Central Engines of NLS1s}
\shortauthors{Ryan et al.}

\begin{document}
      
\title{THE CENTRAL ENGINES OF NARROW-LINE SEYFERT 1 GALAXIES}

\author{C. J. Ryan and M. M. De Robertis\altaffilmark{1}}
\affil{Department of Physics and Astronomy, York University}
\affil{4700 Keele St., Toronto, Ontario, Canada M3J~1P3}
\email{cjryan@yorku.ca, mmdr@yorku.ca}

\author{S. Virani\altaffilmark{1}}
\affil{Department of Astronomy, Yale University}
\affil{P.O. Box 208101, New Haven, CT 06520}
\email{svirani@astro.yale.edu}

\author{A. Laor}
\affil{Physics Department, Technion}
\affil{Haifa 32000, Israel}
\email{laor@physics.technion.ac.il}

\author{and P.C. Dawson} \affil{Physics Department, Trent University}
\affil{1600 West Bank Drive, Peterborough, Ontario, Canada K9J~7B8}
\email{pdawson@trentu.ca}

\altaffiltext{1}{Based on observations obtained at the
Canada-France-Hawaii Telescope (CFHT) which is operated by the
National Research Council of Canada, the Institut National des
Sciences de l'Univers of the Centre National de la Recherche
Scientifique of France, and the University of Hawaii.}

\begin{abstract}

It has been suggested that Narrow-Line Seyfert 1 (NLS1) galaxies are
evolutionarily young objects, powered by the accretion of gas onto
central black holes that are significantly lower in mass than those
found in typical broad-line Seyferts.  We explore this hypothesis
through the analysis of high-spatial resolution, near-IR imaging data
obtained in {\it J} and {\it $K^{\prime}$} for a sample of 11 NLS1s.
Surface brightness profiles are separated into their constituent
components using two-dimensional decomposition techniques.  By
employing the correlation between black-hole mass and host galaxy
bulge luminosity, calibrated for near-IR wavelengths using 2MASS data,
we determine the mean black-hole mass for our sample to be, in solar
units, $\langle{\rm log}(\MBH)\rangle$ = 7.9.  Using the correlation
between the size of the broad-line region and the monochromatic
continuum luminosity, we obtain black-hole mass estimates under the
assumption that the emission-line gas is in virial equilibrium.  The
mean black-hole mass derived from this relation is $\langle{\rm
log}(\MBH)\rangle$ = 6.4.  It is found that the estimates obtained
from the black-hole mass-bulge luminosity relation are systematically
one full order of magnitude larger than those derived from the
black-hole mass-broad-line region radius relation.  We explore
possible causes for this discrepancy in $\MBH$ estimates and the
ramifications for our understanding of the role played by NLS1s in AGN
evolution.

Because numerical simulations constrain the start of the AGN duty
cycle to a time shortly after a significant gravitational interaction,
we examine the morphology and near-IR bulge colors of the NLS1 sample
for evidence of recent encounters.  The mean bulge color is found to
be $\langle({\it J}-{\it K_s})\rangle$ = +1.85 $\pm$ 0.58, which is
redder than that of both a matched sample of non-active galaxies and
published estimates for broad-line Seyferts.  The source of the
unusual bulge colors may be an excess of flux, peaking at around
2.2~\micron, that has been detected near the centers of some NLS1s
such as Mrk 1239.  No evidence is found for light asymmetries or an
extra stellar component that would indicate NLS1s are young objects.
Finally, we postulate that there may be some interesting lines of
circumstantial evidence suggesting that secular processes may be
relevant in the evolution of NLS1s.

\end{abstract}


\section{INTRODUCTION \label{intro}}

The connection between the evolution of galaxies and that of the
supermassive black holes that are known to lurk at their centers is a
key issue in contemporary extragalactic studies.  It is clear that
Active Galactic Nuclei (AGN) represent an important, transient phase
in the development of galaxies at all redshifts \citep{hae93}.  The
source of intense, non-thermal emission generated in the cores of AGN
is widely accepted to be the accretion of material onto otherwise
dormant black holes \citep{ree84, pet00a}.  However we may not yet
fully appreciate the diversity of mechanisms that create the
non-equilibrium conditions leading to accretion.  Significant changes
in gravitational potential can be produced by interactions between
galaxies of comparable mass or by the consumption of dwarf satellite
galaxies \citep{ost93}.  It also may be possible to trigger accretion
through secular processes within the host galaxy such as bar-driven
tidal instabilities, generating episodic bursts of nuclear activity
\citep{kor04}.

An evolutionary link between black holes and their host galaxies is
strongly suggested on the basis of correlations between the black-hole
mass, $\MBH$, and various galaxy parameters.  A good correlation has
been found between $\MBH$ and the central stellar velocity dispersion,
$\sigma_s$, of the spheroid component of the host galaxy
\citep[e.g.,][]{fer00, geb00, gre05a}.  The intrinsic scatter in this
relation may be more significant at low black-hole masses, however
\citep{ben05}.  It has also been shown that $\MBH$ is a constant
fraction of the mass of the host spheroid (or, in the case of
ellipticals, the total mass of the galaxy), with $\MBH$/$M_{sph}
\simeq$ 0.0015 \citep{mag98, mar03, har04}.  Under the assumption of a
constant mass-to-light ({\it M}/{\it L}) ratio, this relation is
frequently expressed in terms of the host bulge luminosity
\citep[e.g.][]{lao01, mcl01}.  While early studies found the scatter
in this mass-luminosity relation to be substantial, recent progress in
the amount and quality of available data has led to an improvement in
the precision of the best-fit parameters \citep{mar03, don06}.
\citet{mar03} also demonstrated the wavelength dependence of the
scatter in the relation, with near-IR observations tracing the mass
distribution better than optical data, generating a tighter
correlation.

There is a general consensus in the literature that AGN exhibit these
correlations independent of distance or luminosity, suggesting that
they are a natural product of galaxy-black hole co-evolution
\citep{kor95, bar05}.  Yet there remains much debate as to how
Narrow-Line Seyfert 1 galaxies, which constitute approximately 15\% of
the total low-redshift Seyfert population \citep{wil02}, fit into this
picture.  NLS1s are a subset of AGN exhibiting an overall spectral
signature consistent with a Type 1 Seyfert, but with remarkably narrow
permitted emission lines.  More specifically, NLS1s are classified
based on an observed broad component of H$\beta$ emission with FWHM
$<$ 2000 km~s$^{-1}$, strong $\Fe$ emission, relatively weak [$\Ox$],
and steeper soft x-ray continuum slopes than broad-line Seyfert 1s
\citep{ost85, bol96}.  Through a Principal Component analysis,
\citet{bor92} found that NLS1s dominate the extreme end of the
``Eigenvector 1'' anti-correlation between the $\Fe$ and [$\Ox$]
emission-line strengths.  This relation is believed to be driven by
the ratio of the luminosity of the source relative to its Eddington
luminosity, $L/L_{Edd}$ \citep{sul00}.  Clearly, NLS1s occupy an
important portion of AGN parameter space.

To account for the extreme properties of NLS1 galaxies, it has been
conjectured that their central engines are black holes of relatively
modest mass (typically of order 10$^6$ {\it M}$_{\sun}$) \citep{bol96,
mat00}.  This hypothesis can explain the narrow emission lines
observed in these objects, under the assumption that the broad
H$\beta$-emitting region is gravitationally dominated by the black
hole.  If the black holes in NLS1s are statistically smaller than in
broad-line Seyfert galaxies, elevated accretion rates would be
required to account for the relatively normal observed luminosities
\citep{pou95}.  Furthermore, it has been suggested that these objects
represent an early stage in AGN evolution, during which the black-hole
mass increases rapidly in a period of intense accretion \citep{mat00}.
As the black hole grows in this scenario, the accretion rate gradually
decreases and the NLS1 matures into a ``normal'' broad-line Seyfert
\citep{gru04a}.

It remains unclear whether the aforementioned correlations with $\MBH$
can be applied to NLS1s.  Several authors suggest that NLS1s fall
below the mass-luminosity correlation of other galaxies, resulting in
smaller estimates of $\MBH$ for a given bulge luminosity
\citep{mat01,wan02}.  It has also been suggested that intense star
formation activity could lead to overestimates of the bulge luminosity
or, equivalently, underestimates of the ({\it M}/{\it L}) ratio
\citep{nel04}.  The validity of the $\MBH$-$\sigma_s$ correlation for
NLS1s has also been questioned.  It has been demonstrated by
\citet{bar05} that AGN with small black holes follow the same relation
as galaxies with more typical values for $\MBH$.  Both \citet{wan01}
and \citet{bot04} estimate the velocity dispersion for NLS1 galaxy
samples from measurements of [$\Ox$], and find that the
$\MBH$-$\sigma_s$ correlation is the same as for broad-line Seyferts.
Conversely, a departure from the standard $\MBH$-$\sigma_s$ relation
for NLS1s was reported by \citet{mat01}, \citet{gru04a}, and
\citet{bia04}.  Each of these studies used [$\Ox$] as a surrogate for
the velocity dispersion.  It is interesting to note that different
conclusions are reached in these studies even though the samples are
similar.

A major thrust of this study is to compare black-hole masses derived
from the well-known correlation between $\MBH$ and the host galaxy
luminosity for the NLS1 subset, hereafter the mass-luminosity
correlation, with other mass estimates.  To this end, we examine a
sample of 11 NLS1s using high-spatial resolution, near-IR imaging
data.  The luminosities of the bulge components in {\it J} and {\it
$K^{\prime}$} are measured independently for each target.  Black-hole
masses are obtained by utilizing the mass-luminosity relation,
calibrated to near-IR wavelengths.  This work builds on previous
studies by applying two-dimensional profile decomposition techniques
to near-IR data of excellent quality, providing an accurate
determination of the photometric properties of the host bulges.  The
results are compared to values of $\MBH$ derived via the correlation
between the radius of the broad-line region and the optical continuum
luminosity, which has been calibrated using reverberation mapping mass
estimates \citep{kas00, kas05}.  This comparison is used to examine
the role of NLS1s in the evolution of AGN.

In \S~2 we outline the sample selection and describe the observations.
The data reduction process, including the decomposition procedure, is
included in \S~3.  The results and a discussion of their implications
for the NLS1 paradigm are presented in \S~4 and \S~5, respectively.
Conclusions are provided in \S~6.  Throughout this work, we adopt
cosmological parameters of $\Omega_{\Lambda}$ = 0.7, $\Omega_m$ = 0.3
and a Hubble constant of H$_{\rm o}$ = 75 km~s$^{-1}$~Mpc$^{-1}$.

\section{OBSERVATIONS \label{obs}}

To examine the possible role of NLS1 galaxies in the evolution of
nuclear activity, a sample was selected from the \citet{ver01a}
catalog, Quasars and Active Galactic Nuclei, 10th Edition.  Targets
were chosen to be low-redshift ($z \leq 0.05$) and moderately bright
(total ${\it K^{\prime}}\leq +12.0$) in order to achieve the
exceptional spatial resolution and signal-to-noise ratios necessary
for accurate profile decompositions.  Selecting targets from the
\citeauthor{ver01a} catalog also ensured that a reasonable amount of
complementary observations would be available, including optical
spectroscopic data \citep{ver01b}.  A total of 150 objects are
classified as NLS1s in the \citeauthor{ver01a} catalog.  Of these,
approximately 37 matching our selection criteria are visible in the
Northern sky.  We imaged 11 of these NLS1s in December 2002 through
the broad-band {\it J} and {\it $K^{\prime}$} filters at the 3.6-meter
CFHT using KIR, the $1024 \times 1024$ HgCdTe array and near-IR
camera, mounted at the f/20 output focus of the PUEO Adaptive Optics
Bonnette \citep{rig98}.  The focal plane scale of this system is
0.0348\arcsec /pixel, providing a total field of view of
36\arcsec~$\times$ 36\arcsec.  For all of the target observations, the
galaxy nucleus was sufficiently bright to be used as the wavefront
reference source.  The photometric standard stars were also observed
in the same way, acting as their own wavefront reference sources.
This greatly facilitated the registration and co-addition of all
dithered sequences, rendering other corrections unnecessary.  Strehl
ratios for the galaxy observations were estimated to be between 0.2
and 0.4, while for the standard stars the Strehl ratio was found to be
approximately 0.5 \citep{beu98, rig98}.  A PSF FWHM of 0.2\arcsec~was
consistently measured from the standard star exposures, providing
significant spatial over-sampling of the data.  The key photometric
properties of the galaxies, including morphological type, optical
semi-major axis, and {\it K} = +20 mag arcsec$^{-2}$ isophotal radius,
are provided in Table~\ref{params}.  We note that the {\it K}-band
isophotal radii of the targets are between 6.4\arcsec~and 26.1\arcsec.

An exposure sequence for each object consisted of six
spatially-dithered images, including two off-target exposures that
were subsequently used to carefully monitor the near-IR sky and to
provide accurate background subtraction.  The data were acquired using
the sequence T, T, B, T, T, B, with T representing an on-target
exposure and B an off-target background exposure.  The off-target
exposures were dithered by between 1.6\arcmin~and 2.3\arcmin,
depending on the size of the galaxy, to the East and West sides of the
on-target field of view in order to avoid contamination by the
extended disks of the galaxies.  This procedure ensured that the
on-target images were bracketed, both spatially and temporally, by the
background frames.  The on-target exposures were typically dithered by
approximately 16\arcsec~in a four-point box pattern that placed the
galaxy nucleus in each of the detector's quadrants.  Total on-target
exposure times varied between 80 and 1920 seconds, chosen to maximize
the signal while ensuring that the detector response remained linear.
Equal exposure times were used for all images in each sequence.  The
mean airmass for each data set was typically between 1.0 and 1.2.  A
summary of the observation log is provided in
Table~\ref{observations}.

\section{DATA REDUCTION \label{data}}

The data were reduced using standard IRAF procedures, including the
construction and application of bad pixel masks, the removal of dark
current, and a flat-field correction (including any large-scale
gradients).  A crucial step in the reduction and analysis of near-IR
images is the accurate subtraction of the sky background.  The
characterization of the background level and pattern is particularly
challenging, as it has been shown that the mean level at the CFHT site
can vary by 0.7\% and 0.5\% in {\it J} and {\it $K^{\prime}$},
respectively, on time scales of one minute \citep{vad04}.  Failure to
take this into consideration could potentially lead to photometric
errors when imaging extended sources with lengthy integration times.
Because of the relatively large angular size of the targets compared
with the detector's field of view, it was impractical to determine the
background level using adjacent on-target exposures.  Instead,
accurate background subtraction was achieved by carefully monitoring
the two off-target exposures from each data set.  An inspection of the
median and mode for each pair of off-target frames indicated that the
background level and underlying patterns were constant to within 1\%,
allowing us to use the combined off-target frames for background
subtraction.

We then examined the on-target exposures, both before and after
background subtraction, to search for any systematic uncertainties
introduced by this procedure.  As the on-target dither pattern
situated the galaxy nucleus sequentially in each of the detector's
four quadrants, we monitored the image statistics of the opposite
quadrant carefully.  Variations in the median count rate among the
image corners were always within underlying statistical uncertainties.
The residual flux levels of the corners of the background-subtracted
images were consistently near zero, with no residual gradients
present.  This gave us confidence that the background had been
successfully and appropriately removed in all target exposures.

The dithered on-target exposures were combined into 56\arcsec~$\times$
56\arcsec~mosaic images using the IRAF package {\it irred.irmosaic}.
For three galaxies, multiple image sets were obtained in {\it
$K^{\prime}$}, allowing independent mosaics to be generated.  This
served as a useful check on the consistency of the data reduction and
analysis processes.

\subsection{Bulge-Disk Decomposition \label{decomp}}

The surface brightness profiles of the reduced mosaic images were then
examined to determine the relative flux associated with the bulge
component.  To accomplish this, the two-dimensional profile fitting
algorithm GALFIT v2.0 was employed.  GALFIT allows the user to model
the light profile of a target using a suite of analytic functions,
including Gaussian, exponential, S\'{e}rsic/de Vaucouleurs, and Nuker
profiles.  The optimal fit is determined, based on the user's
selection of input parameters, by minimizing the $\chi^2$ residual
between original and model profiles.  Details of the minimization
routine are provided in \citet{pen02}.

While GALFIT is capable of fitting a PSF-convolved Gaussian to account
for a point source, in the interest of limiting the number of free
parameters in the model and to best ensure the parameters are
physically realistic, an alternative method was used to subtract the
contribution of the active nucleus to the integrated flux.  This {\it
shift-scale-subtract} (sss) routine involves using a high
signal-to-noise ratio, isolated point source such as a photometric
standard star observed under similar conditions to model the AGN
\citep{vir00}.  For each target, a standard star observed close in
time was selected as a template point source.  The stars were imaged
with the AOB activated, using the same dither pattern as employed for
the target observations.  The spatial and temporal stability of the
standard stars was monitored by examining the reduced exposures.  It
was found that the stellar PSFs were consistent to within 0.02
magnitudes, and the FWHM varied by less than 0.05\arcsec.  To subtract
the nuclear component, the star was aligned to the center of the
galaxy as estimated from elliptical isophote fitting via the {\it
ellipse} package.  Accuracies of $\pm 0.1$~pixel are achievable for
the spatial alignment by monitoring asymmetries in the light profile
of the difference image.  The scaling of the nuclear PSF was judged to
be that which produced a nearly flat profile in the central
0.4\arcsec.  The average contribution of the nuclei to the total
galaxy flux was determined to be approximately 15\% in {\it J} and 9\%
in {\it $K^{\prime}$}.  It was thus possible to assess the nuclear
contribution to better than 1\% of the total galaxy luminosity.
Appropriate nuclear scale factors were determined independently for
the {\it J}- and {\it $K^{\prime}$}-band images.

The nucleus-subtracted galaxy images were then modeled as the sum of a
bulge and, when necessary, disk component using GALFIT.  Galaxy disks
were fit using an exponential profile \citep{fre70}.  The analytic
form of galaxy bulges is less clear, and has been discussed
extensively in the literature.  Most early studies represented bulge
light as a \citet{dev48} $r^{1/4}$ profile, similar to that used for
elliptical galaxies.  However, for many galaxies the profile fails to
accurately reproduce the shape of the bulge at all radii
\citep{cao93}.  This departure from the $r^{1/4}$ profile is of
greatest concern for early-type galaxies, where a significant fraction
of the total galaxy light comes from the bulge component.  It has been
demonstrated that bulges are better described using the more general
\citet{ser68} $r^{1/n}$ profile \citep{and95}, which reduces to a de
Vaucouleurs profile for {\it n} = 4.  S\'{e}rsic profiles were applied
to model the bulge light for the NLS1 sample with the index, {\it n},
a free parameter to be determined by GALFIT.  As all of the targets
are early-type galaxies, no additional components were required to
account for features such as spiral arms.

Initial estimates for the disk scale radii were taken from the 2MASS
Extended Source Catalog (XSC), recalibrated to the KIR image scale.
The XSC images have a relatively coarse 1\arcsec /pixel focal plane
scale and an integration time of 7.8 seconds.  As such, the published
scale parameters for these moderately faint objects, which have a mean
2MASS {\it $K_s$} = +10.8, were treated as initial estimates to be
determined more accurately by GALFIT.

Starting values for the bulge scale radii were then estimated from an
examination of disk-subtracted, one-dimensional surface brightness
profiles.  The position angle, ellipticity and {\it a4} ``diskiness
parameter'' were also estimated from elliptical isophote fitting to
the nucleus-subtracted images.  For the S\'{e}rsic index, an initial
estimate of {\it n} = 4 was used, as the targets are generally
early-type galaxies and it has been demonstrated that there is a trend
of decreasing index as morphological type proceeds from early to late
\citep{gra01a}.  All of these parameters were allowed to vary within
the $\chi^2$ minimization process, however.  The luminosity
bulge-to-disk ratio was estimated from the empirical relation with
Hubble stage originally presented in \citet{sim86} and derived from
{\it K}-band data in \citet{gra01a}.

For each target, we used GALFIT to first obtain one-component fits to
the disks of the galaxies.  Pixel masks aligned to the isophotal
centers of the galaxies were used to block the bulge components in
these initial model fits.  The results were then used as estimates of
the input disk parameters in two-component fits, again allowing all
parameters to vary in order to find the global $\chi^2$ minimum.  For
these fits, we did not apply pixel masks to any sections of the mosaic
images and fit the galaxy profiles to {\it r} = 0.

It is also necessary that the relevant parameters be physically
realistic.  For three targets, Mrk 359, Mrk 1044 and MCG08.15.009, it
was necessary to constrain one of the free parameters in order to
avoid convergence to a non-physical analytic fit.  For these galaxies,
the bulge scale radius was held fixed to the value provided by the
2MASS XSC.  Residual images were also visually examined for any
significant features or asymmetries.

Proper background subtraction is essential to measure accurate
integrated fluxes for each component of the galaxy.  To estimate the
uncertainties in the fitted parameters to reasonable uncertainties in
the background level, we performed fits to the data with sky levels
set systematically high and low by one standard deviation in the
background.  The same input parameters used to derive the optimal fits
were employed in these tests.  This technique is similar to that
applied to one-dimensional decompositions by \citet{gra01a}.  In some
cases, negative fluxes introduced by background over-subtraction
prevented GALFIT from converging on a meaningful solution.  Based on
the results of these extensive tests, however, it was established that
conservative uncertainties in the bulge fits resulting from imperfect
background subtraction are less than 0.1 magnitudes.

\begin{figure}
\plotone{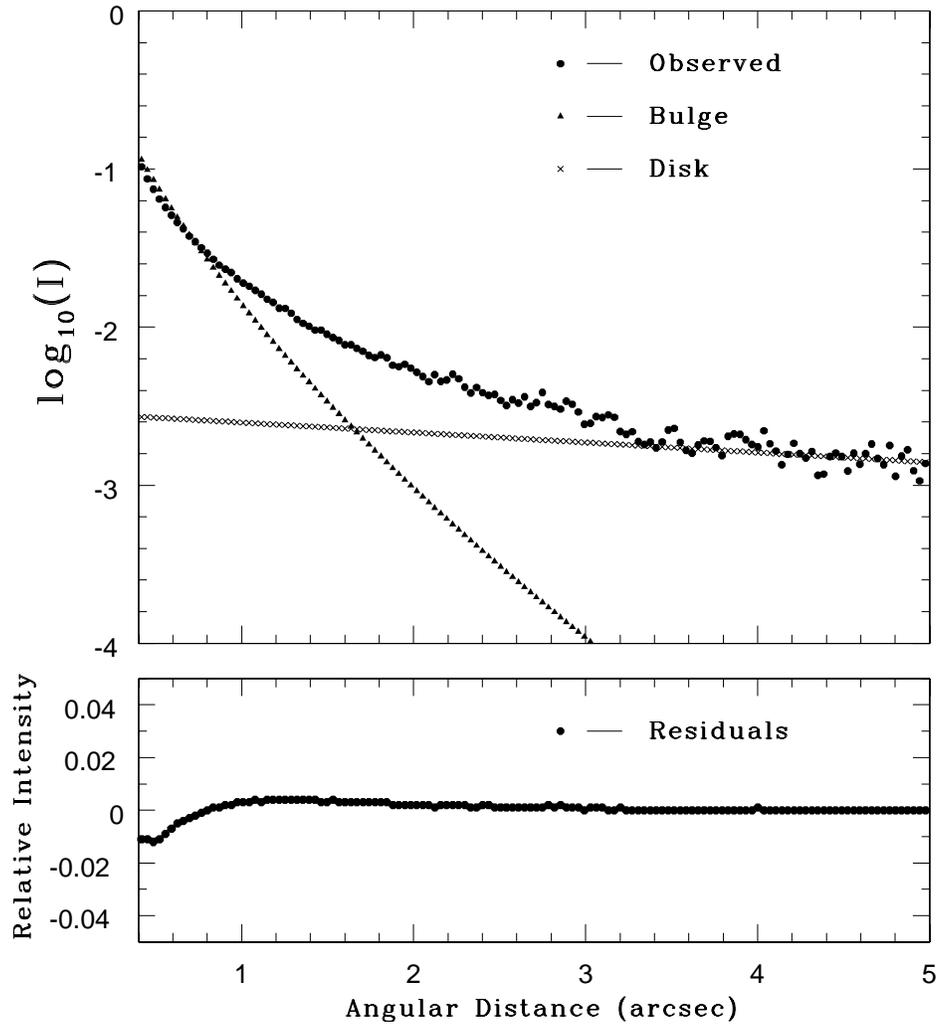} \figcaption{{\it J}-band surface brightness profile
for Mrk 1239.  The top panel shows the relative contribution of each
component.  In the lower panel, the residuals (observation minus
model) are plotted.
\label{sbprofiles}}
\end{figure}

\begin{figure}
\plotone{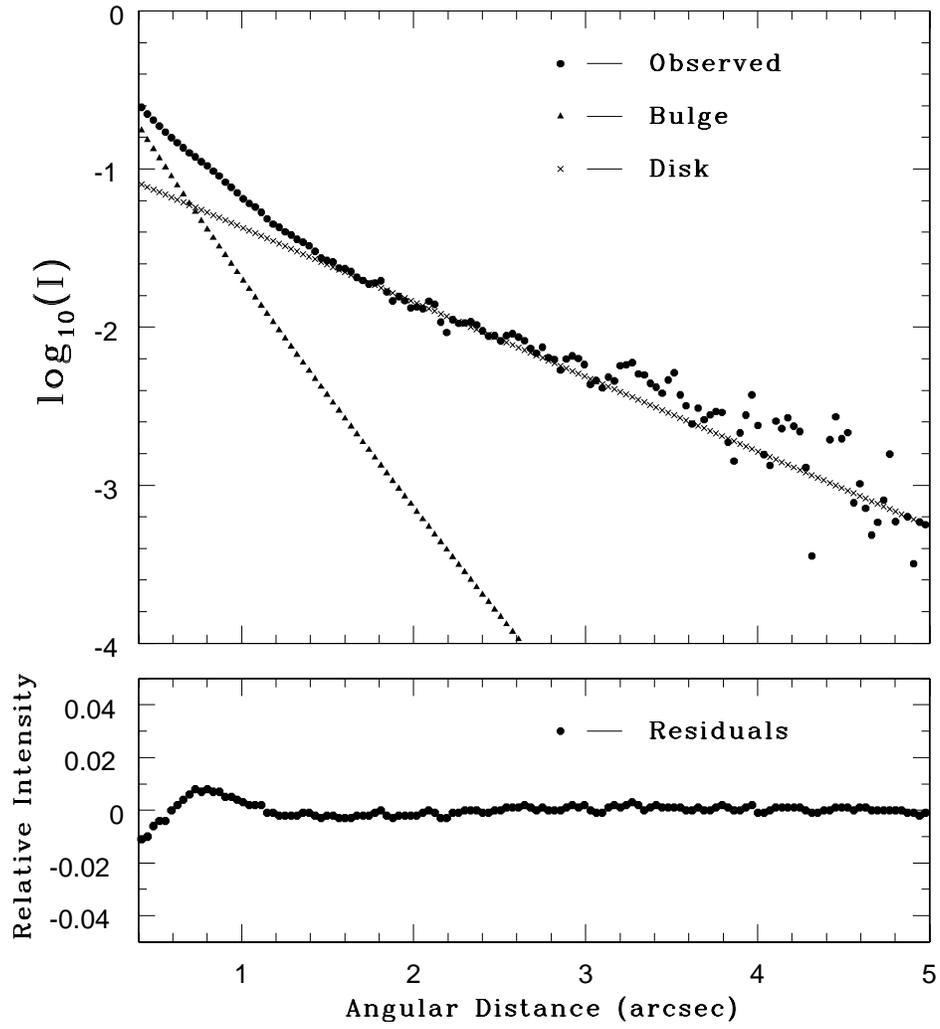} \figcaption{{\it J}-band surface brightness profile
for IRAS 04596-2257.
\label{sbprofiles2}}
\end{figure}

\begin{figure}
\plotone{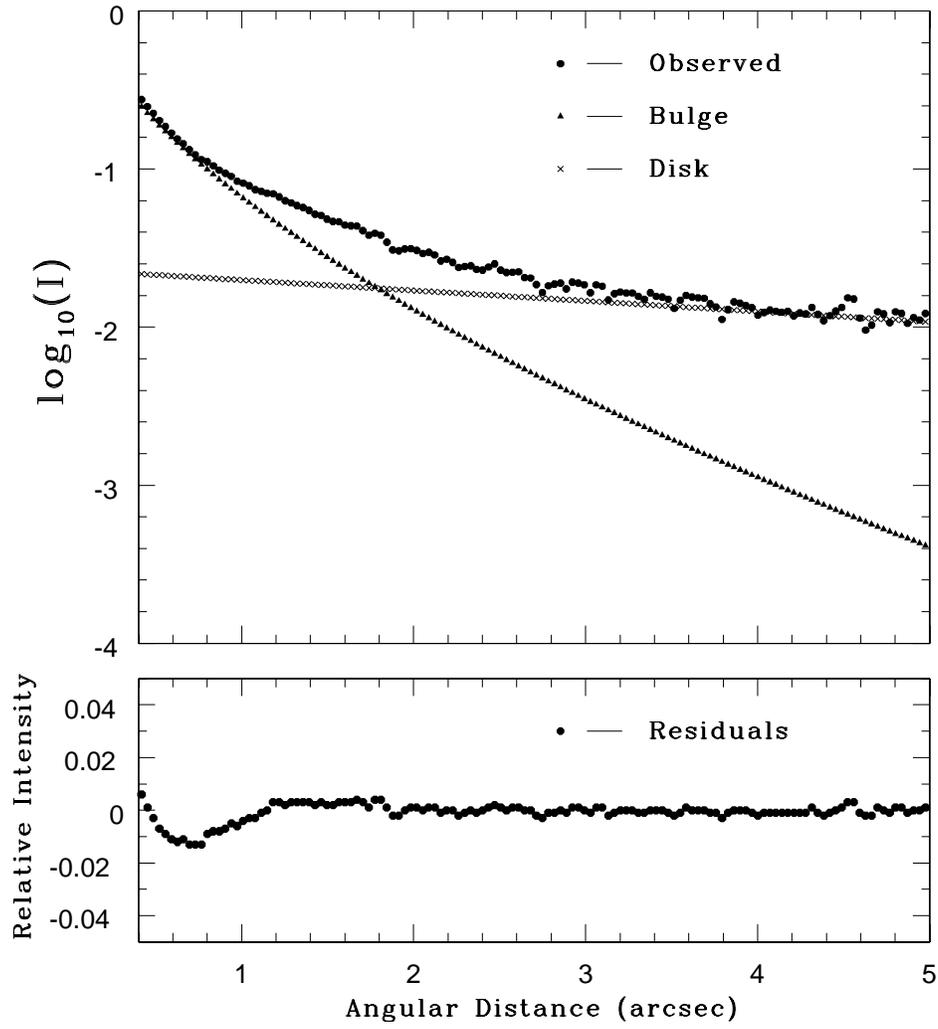} 
\figcaption{{\it K$_s$}-band surface brightness profile for Mrk 1126,
illustrating the residual pattern seen near the centers of many of the
model fits in {\it K$_s$}.
\label{kprofile}}
\end{figure}

Separate model fits to the {\it J} and {\it $K^{\prime}$} images were
obtained for each galaxy.  Plots of the {\it J}-band surface
brightness decompositions are shown in Figures~\ref{sbprofiles}
through~\ref{sbprofiles2}.  For each target, the relative contribution
of the components are presented in log space; residuals to the fit are
plotted separately in linear space.  For the {\it $K^{\prime}$}-band
decompositions, we find that there is a consistent pattern in many of
the residual profiles on angular scales less than 1\arcsec~from the
galaxy center, such that a broad negative peak is followed by a small
upturn.  This is illustrated in Figure~\ref{kprofile}, which presents
the {\it $K^{\prime}$}-band profile for Mrk 1126.  It is conceivable
that the pattern, which has an insignificant effect on total bulge
magnitudes, may be due to limitations in our nucleus-subtraction
routine.  In order to search for any systematic effects introduced by
the sss-technique, we performed profile decompositions on images in
which the point sources were purposely over-subtracted and
under-subtracted by 10\%.  This level of uncertainty is greater than
could reasonably be expected from a careful subtraction of the
nucleus.  We found in these cases that the bulge magnitudes varied by
of order $\pm 0.03$~magnitudes.  The reason this has no significance
on total bulge magnitudes is essentially because the angular extent of
the model bulge profiles are between 2\arcsec~and 8\arcsec.  Since
nuclei affect only the central 0.2\arcsec, their influence is limited
to a fraction of the inner 1\% of the two-dimensional bulge profiles.

In employing the sss-technique, it is assumed that the PSFs associated
with the active nuclei can be adequately modeled as scaled versions of
the photometric standard stars.  However, small discrepancies could
arise from effects such as changes in the airmass of the observations
or contamination from the AGN host galaxies.  Furthermore, differences
in the wavefront reference source magnitudes between the AGN and
standard star observations resulted in small variations in the Strehl
ratios.  This could produce slight differences in the AO-corrected
PSFs.  To further verify that the subtraction of the flux associated
with the active nucleus did not introduce any systematic effects on
the bulge fits, we performed three-component fits to the data using
GALFIT.  In addition to the bulge and disk components, a nuclear point
source was included, represented by a Gaussian function.  The bulge
magnitudes and S\'{e}rsic indices derived from these fits were
consistent with the results obtained using the sss-technique.  The
quality of the fits, as parameterized by the reduced $\chi^2$ values,
were also comparable.  Additionally, two-component (bulge+disk) fits
were performed on the non-PSF-subtracted images.  The output bulge
magnitudes were again similar to the results presented in
Table~\ref{galfitparams}, albeit with slightly higher S\'{e}rsic
indices and increased residuals.  These results are not surprising,
given that the flux associated with the nucleus is of order a few
percent of the total galaxy flux, and is confined to the central 1\%
of the model fits.  From these tests it was concluded that the
sss-technique is robust, and does not compromise the model bulge
parameters.

The small but systematic {\it $K^{\prime}$}-band residuals may also
reflect a departure of the analytical model from the physical
structure near the center of the bulge component that is resolved by
our data.  An additional source of emission may influence the surface
brightness on scales less than 1\arcsec~from the nucleus
\citep{rod06}.  This option is discussed in greater detail in \S~5.
The net effect of these small, systematic residuals on the total
integrated luminosities of the galaxy components is negligible.

For the three galaxies that were observed twice in {\it $K^{\prime}$},
independent fits provided a useful check of the consistency of the
fitting procedure.  We found that the bulge magnitudes obtained from
these images were consistent within the formal uncertainties.
Absolute bulge magnitudes were then calculated using redshift
information taken from the literature and Galactic extinction
corrections based on the dust maps presented in \citet{sch98}.  The
results are summarized in Table~\ref{galfitparams}.

\section{BLACK-HOLE MASSES \label{masses}}

Black-hole mass estimates for AGN are frequently obtained by employing
the relation between the size of the broad-line region and the
monochromatic continuum luminosity at 5100\,\AA, the mass-radius
relation.  We will compare these results with masses derived from the
well known correlation between $\MBH$ and the host bulge luminosity,
the mass-luminosity relation, using the results obtained from the
two-dimensional profile decompositions.  In the following subsections
we discuss these techniques in greater detail, and present the
resulting black-hole mass estimates.

\subsection{The Mass-Radius Relation \label{massrad}}

Reverberation mapping is widely considered to be a reliable method to
obtain accurate $\MBH$ estimates in Seyfert 1 galaxies, for which
direct probing of the black-hole sphere of influence is not possible
due to the intense nuclear emission.  Correlations with host galaxy
parameters derived from reverberation mapping results are generally
consistent with those of non-active galaxies with secure values of
$\MBH$ \citep{fer05}.  However direct measurements of the time delay
between fluctuations in continuum and line emission can be challenging
and require long-term monitoring of the source.  As a consequence,
estimates of $\MBH$ are more commonly generated using secondary,
indirect methods such as exploiting the correlation between the radius
of the broad-line region and the monochromatic continuum luminosity at
optical wavelengths \citep{kas05}:
\begin{equation}\label{kaspi}
\frac{R_{BLR}}{10\,\rm{lt-days}} = (2.23 \pm 0.21) \left[\frac{\lambda L_\lambda(5100\,{\rm{\AA}})}{10^{44}\,\rm{erg\,s}^{-1}}\right]^{0.69 \pm 0.05}.
\end{equation}

Based on a sample of six objects with FWHM(H$\beta$) $<$ 2000
km~s$^{-1}$, \citet{pet00b} suggested that this relation can be
extended to the NLS1 subset.  This conclusion is supported by the
observation that the continuum luminosity at 5100 \AA~is comparable in
NLS1s and broad-line Seyferts \citep{gru04b}.  However, the low
optical variability that is characteristic of NLS1s makes secure
reverberation mapping masses of these objects particularly difficult
to obtain, as demonstrated by the small number that have been
successfully monitored \citep[e.g.][]{she01}.

Optical spectra were published by \citet{ver01b} for a large number of
NLS1s, including seven of the objects in our sample.  The data were
obtained with the CARELEC spectrograph mounted to the 1.93-meter
telescope at the Observatoire de Haute-Provence, providing a
resolution of 200 km s$^{-1}$ at H$\beta$.  The contribution of $\Fe$
emission was removed using a template spectrum generated from
observations of the NLS1 I Zw 1, according to the procedure discussed
in \citet{bor92}.  The published spectra have a wavelength range of
4750 to 5100 \AA.  The paper also includes values for the FWHM of the
broad component of H$\beta$ for each object, taken from the
literature.  We measured the monochromatic continuum luminosity at
5100 \AA~from these spectra, correcting for Galactic extinction.  The
results were used to estimate the radius of the broad-line region
according to (\ref{kaspi}).  Assuming the emission-line gas is in
virial equilibrium with the central black hole, estimates of $\MBH$
for each object may then be obtained using the following expression:
\begin{equation}\label{virial}
M_{BH} = R_{BLR}\,G^{-1}(kv_{H\beta})^2.
\end{equation}
The constant {\it k} is a function of the geometry of the broad-line
region.  Assuming the line-emitting gas is distributed purely
isotropically, as \citet{kas00} and \citet{wan02}, {\it k} =
$\sqrt{3}/2$.

For each of our targets with spectroscopic data available in
\citet{ver01b}, values for the continuum luminosity at 5100 \AA, the
FWHM of the broad component of H$\beta$, and the calculated black-hole
masses are included in Table~\ref{bhmasses}.  The mean value of $\MBH$
for the seven NLS1s is $\langle{\rm log}(\MBH)\rangle$ = 6.3, in
agreement with the idea that black holes in NLS1s are statistically
less massive than broad-line Seyferts.  The assumption of a
spherically symmetric gas distribution, however, is likely to be an
oversimplification.  A physically more meaningful model may include a
gas disk in addition to an isotropic component \citep{mcl01}.  The
introduction of a disk component in the broad-line region would
increase the virial mass estimates presented here by a factor of
three.

\citet{gre05b} propose a correlation between the radius of the
broad-line region and Balmer emission-line luminosities.  Using a
sample of 229 objects taken from the Sloan Digital Sky Survey
(\citealt{yor00}), they obtain an empirical relation between $\MBH$
and the H$\beta$ line luminosity of the form:
\begin{equation}\label{greene}
M_{BH} = (3.6 \pm 0.2)\times 10^6 \left[\frac{L_{H\beta}}{10^{42}\,\rm{erg\,s}^{-1}}\right]^{0.56 \pm 0.02}
\left[\frac{\rm FWHM_{H\beta}} {10^3\,{\rm km\,s}^{-1}}\right]^2 M_\sun.
\end{equation}
In formulating this relation, the authors derive the mass-luminosity
relation independent of \citet{kas05}.  This provides a second method
to estimate masses based on reverberation mapping.

We measured the H$\beta$ line flux for the seven NLS1s from our sample
with available optical spectra and used the results to compute the
black-hole mass in each object according to (\ref{greene}).  The
values obtained for L(H$\beta$) and the resulting estimates of $\MBH$
are included in Table~\ref{bhmasses}.  The average unweighted
black-hole mass was found to be $\langle{\rm log}(\MBH)\rangle$ = 6.4,
again consistent with the notion that black holes in NLS1s are
statistically less massive than those found in broad-line Seyferts.

\subsection{The Mass-Luminosity Relation \label{masslum}}

As discussed in \S~1, it is generally accepted that the slope of the
correlation between $\MBH$ and host bulge luminosity is consistent for
both non-active galaxies and ``typical'' Seyfert galaxies.  However,
it remains a matter of debate as to whether the relation can be
extended in the same form to the NLS1 subset.  Using the results
obtained from our profile decompositions, we now derive estimates of
$\MBH$ under the assumption that the slope of the correlation is the
same for all Seyfert galaxies, including NLS1s.  We then compare these
mass estimates to those obtained in \S~4.1.  The results are presented
in this manner in order to avoid {\it a priori} assumptions about the
accuracy of the black-hole masses obtained from the mass-radius
relation.  While reverberation mapping masses are generally secure,
only a handful objects that could be classified as NLS1s have been
directly observed using this technique.  It is conceivable that other
factors such as the orientation of the broad-line region could affect
the observed line widths, potentially leading to underestimates of the
black-hole masses.  Such issues do not affect the mass-bulge
luminosity relation.

The correlation between $\MBH$ and the mass of the spheroid component
of the host galaxy is frequently quantified in terms of the {\it
B}-band magnitude \citep[e.g.][]{fer05}.  However, the blue luminosity
is biased by the presence of young, hot stars and does not necessarily
best represent the underlying mass distribution.  Near-IR wavelengths
are less sensitive to this bias, and are less affected by dust
extinction.  This is evident from a comparison of the mean extinction
curves in {\it B} and {\it $K^{\prime}$}.  Using values presented in
\citet{mat90}, the ratio of {\it B}- to {\it $K^{\prime}$}-band
optical depths is approximately 12.3.  Based on the extinction values
for our target fields, we estimate that the attenuation of light in
{\it B} due to the presence of Galactic dust is more than 11 times
greater than that in {\it $K^{\prime}$}.  Hence the scatter in the
correlation can be reduced by employing near-IR data, as first
demonstrated by \citet{mar03}.

We calibrated the mass-luminosity relation for near-IR passbands using
a sample of 16 bright elliptical galaxies with secure mass estimates
obtained by resolving the black-hole sphere of influence as reported
by \citet{fer05}.  The assumption that ellipticals are kinematically
similar to the bulge components of spiral galaxies and therefore
follow the same correlation is supported by observational evidence for
consistent Fundamental Plane relationships \citep{kor82, fal02}.
Excluding spiral galaxies from the sample eliminates the need for
separating the total integrated flux into bulge and disk components,
thus minimizing the scatter in the correlation.  Furthermore, we
reject several ellipticals with secure $\MBH$ estimates for the
following reasons.  Cygnus A is known to be a member of a rich cluster
of galaxies, and thus may be influenced by a complex gravitational
potential.  Two other galaxies, NGC 821 and NGC 2778, have estimated
spheres of influence that are smaller than the spatial resolution of
the data.  For each remaining object, the total magnitude in the 2MASS
near-IR bandpasses of {\it J} and {\it K$_s$} were extracted from the
XSC.  It is important to note that the response curves of the 2MASS
filter set are not identical to those of the CFHT filters used for our
NLS1 observations.  The necessary photometric transformations are now
discussed.

Absolute magnitudes were calculated for all but two of the galaxies
using distance estimates from \citet{ton01}.  As data were not
available in \citeauthor{ton01} for NGC 6251 and NGC 7052, we
determined luminosity distances using redshifts obtained from the
Third Reference Catalog of Bright Galaxies (RC3)\citep{dev91}.  A
summary of the properties of the galaxies used to calibrate the
mass-luminosity relation is presented in Table~\ref{ellipticals}.
 
\begin{figure}
\plotone{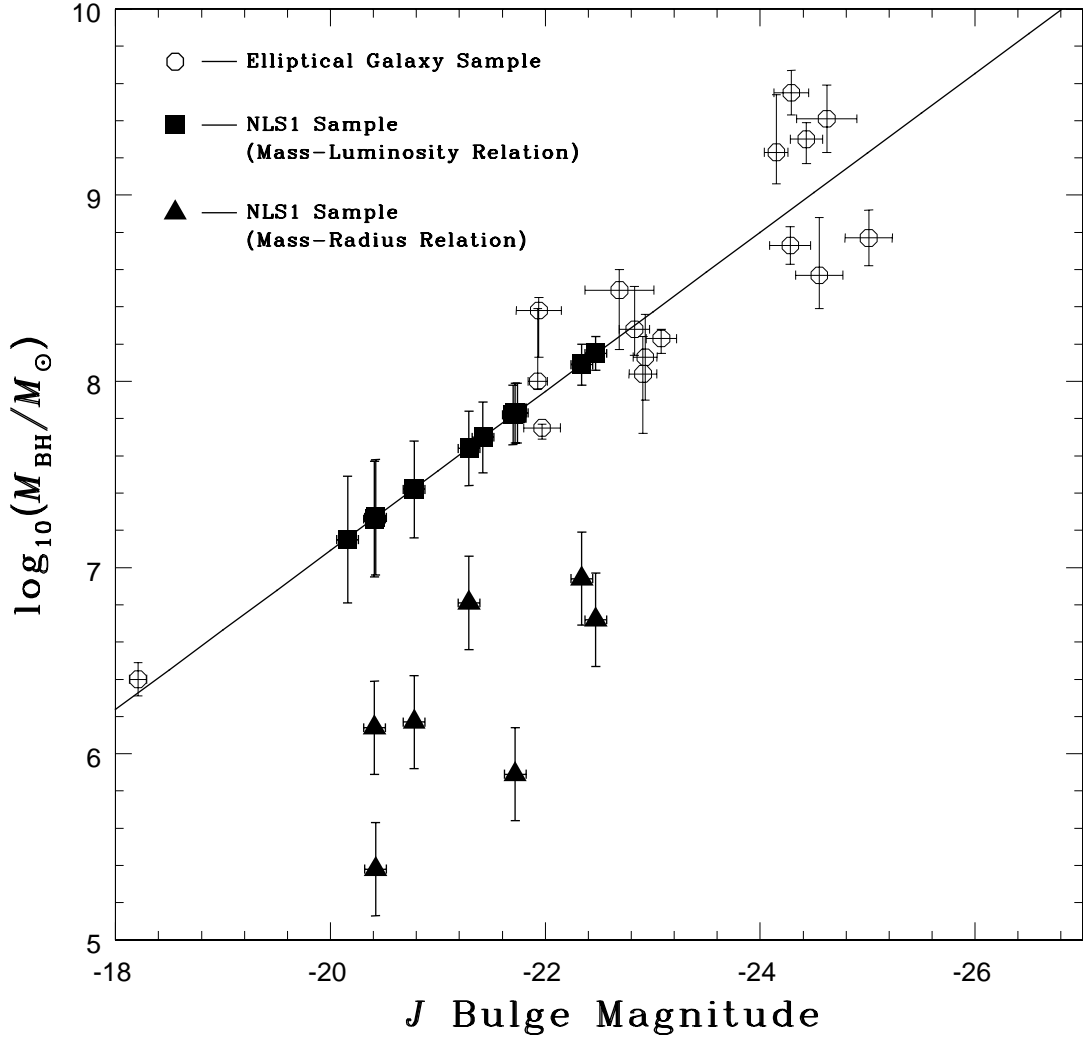}
% Note to editor: for electronic edition, please use the following
% command to include color version of this figure.
%\plotone{f13_colorversion.eps}
\figcaption{Correlation between black hole mass and host galaxy bulge
luminosity, calibrated for {\it J}-band data.  Open circles (colored
blue in electronic edition) denote elliptical galaxies with secure
estimates of $\MBH$.  NLS1 mass estimates computed using this
correlation are plotted as (red) squares.  NLS1 black-hole masses
derived using the mass-radius relation are denoted by (green)
triangles.  The (blue) solid line represents the weighted linear
least-squares fit to the elliptical galaxy data.
\label{bh_bulge_j}}
\end{figure}

\begin{figure}
\plotone{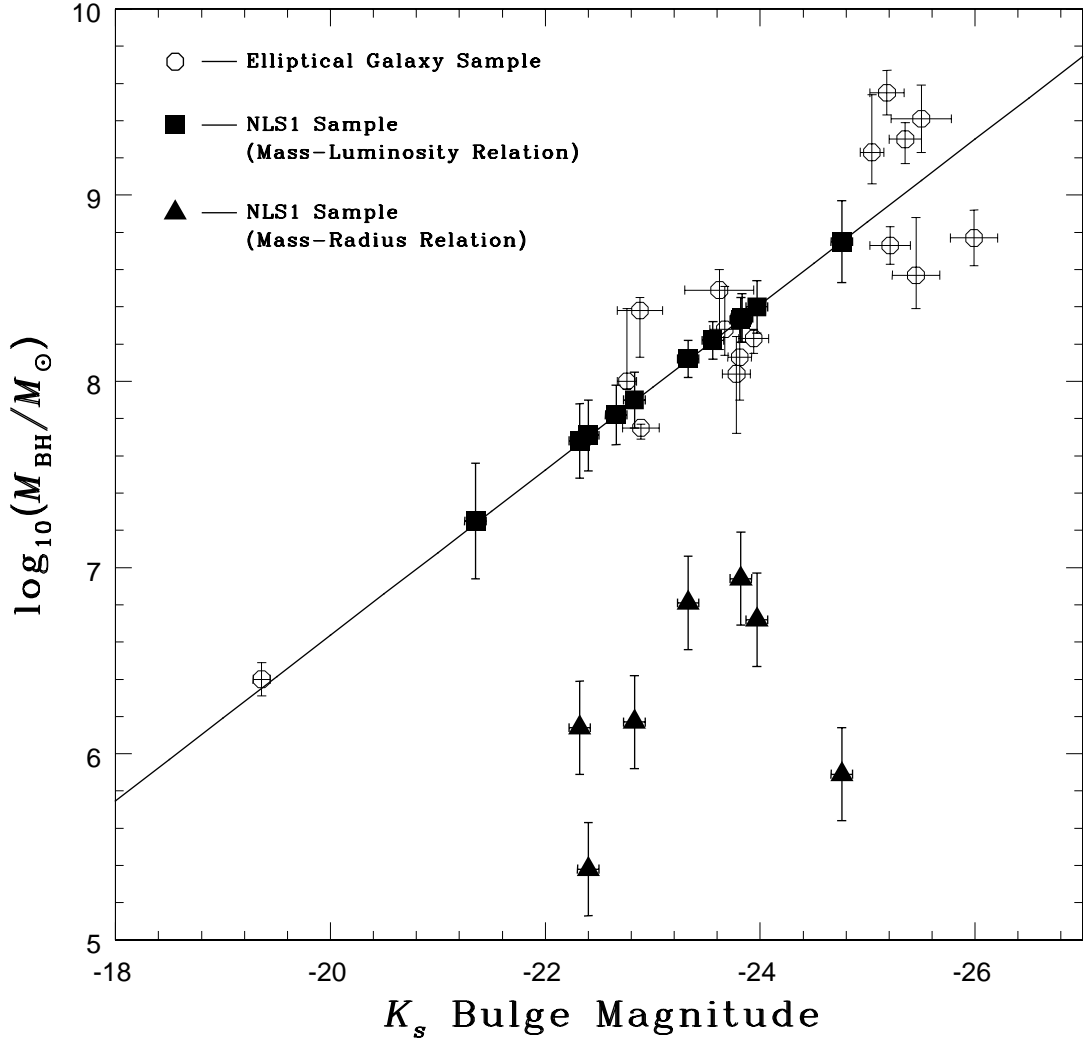} 
% Note to editor: for electronic edition, please use the following
% command to include color version of this figure.
%\plotone{f14_colorversion.eps}
\figcaption{Correlation between black hole mass and host bulge
luminosity, calibrated for {\it K$_s$}-band data.
\label{bh_bulge_k}}
\end{figure}

In Figures~\ref{bh_bulge_j} and \ref{bh_bulge_k}, we plot the
logarithm of $\MBH$, in solar masses, versus the absolute magnitude in
{\it J} and {\it K$_s$} for the sample of elliptical galaxies.  The
best-fit slopes to the data were determined from a linear
least-squares fit, weighting the parameters by their uncertainties
\citep{pre92}.  The mass-luminosity correlations in {\it J} and {\it
K$_s$} have the following forms:
\begin{equation}\label{jband}
{\rm log}\,\MBH = (-0.43 \pm 0.04)(M_J + 22.52) + (8.17 \pm 0.09)
\end{equation}
\begin{equation}\label{kband}
{\rm log}\,\MBH = (-0.44 \pm 0.04)(M_{K_s} + 23.45) + (8.17 \pm 0.09).
\end{equation}
Our results are consistent within uncertainties with both the {\it
K$_s$}-band fit presented in \citet{don06}, obtained using the same
sample criteria, and the results of \citet{mar03}, whose sample
includes spiral galaxies.

In order to use these correlations, it was necessary to convert the
bulge magnitudes obtained by GALFIT from the Mauna Kea Observatories
(MKO) photometric system to the 2MASS system.  Color transformations
for various filter sets are available through the 2MASS
website\footnote{http://www.ipac.caltech.edu/2mass/releases/allsky/doc/sec6\_4b.html}.
The transformations for the MKO system are:
\begin{equation}
K_{s,\,2MASS} = K_{MKO} + (0.002 \pm 0.004) + (0.026 \pm 0.006)(J - K)_{MKO}
\end{equation}
\begin{equation}
(J - K_s)_{2MASS} = (1.037 \pm 0.009)(J - K)_{MKO} + (-0.001 \pm 0.006).
\end{equation}
However, these transformations are valid for the {\it K} filter and
not {\it $K^{\prime}$}.  \citet{wai92} compare the {\it $K^{\prime}$}
filter to {\it K}, and suggest a color transformation of the form
\begin{equation}
K^{\prime} - K = (0.22 \pm 0.03)(H - K).
\end{equation}

In the absence of {\it H}-band data for our sample galaxies, we
adopted a mean ({\it H}-{\it K}) color estimated from \citet{fio99}.
Using a sample of approximately 1000 galaxies, the authors computed
near-IR colors as a function of host morphology.  They determine a
median ({\it H}-{\it K}) for ellipticals of approximately 0.20 $\pm$
0.01, rising to 0.27 $\pm$ 0.01 for spirals.  Assuming that the bulge
colors for our galaxies are similar to that of ellipticals, a
conversion from {\it $K^{\prime}$} to {\it K} is possible.  Combining
this result with the color transformation between {\it K} and {\it
K$_s$}, it can be seen that magnitudes measured in {\it $K^{\prime}$}
and {\it K$_s$} are the same within photometric errors.  This is
supported by an inspection of the transmission functions for {\it
$K^{\prime}$} and {\it K$_s$} filter, which are very similar
\citep{kim05}.  A non-trivial transformation between {\it J$_{MKO}$}
and {\it J$_{2MASS}$} is required of course.

It was also necessary to adjust our bulge magnitudes by applying
modest {\it K}-corrections to our data.  Corrections were taken from
\citet{pog97} who used evolutionary synthesis models based on redshift
and as a function of host morphology.  When necessary, corrections
were estimated from a linear interpolation between redshift bins of
width 0.02.  The measured bulge magnitudes were then inserted into
(\ref{jband}) and (\ref{kband}) to estimate $\MBH$ for each NLS1
target.  The results are presented in Table~\ref{bhmasses}.  The
unweighted mean mass was found to be $\langle{\rm log}(\MBH)\rangle$ =
7.9.

As a consistency check, we applied the \citet{sim86} relationship
between the luminosity bulge-to-disk ratio and Hubble stage.  This
method is less reliable than our more accurate component decomposition
due to the large dispersion in the relation.  Using a sample of 118
spiral galaxies, representing morphological stages S0/a through Sc,
\citet{don06} have calibrated the relation for the 2MASS {\it
K$_s$}-band.  The relation is represented by the following cubic
equation:
\begin{equation}\label{morph}
\Delta m_{K^\prime} = 0.297(T + 5) - 0.04(T + 5)^2 + 0.0035(T + 5)^3.
\end{equation}
Of the 11 NLS1s included in our sample, published morphological
classes were available for seven galaxies.  Preliminary
classifications were made for two additional galaxies, MCG 08.15.009
(T=$-1$) and PG 1016+336 (T=1), based on a visual inspection of our
high-spatial resolution data.  Morphological information was not
available for Mrk 734 and IRAS 04596-2257.  Applying (\ref{morph}) to
the sample, bulge magnitudes were calculated and compared to those
derived by GALFIT.  The results are included in Table~\ref{bhmasses}.
While the scatter in the measurements for individual galaxies is
substantial, the mean bulge magnitudes are consistent to within 0.02
magnitudes, further establishing that the values generated by GALFIT
do not suffer from any systematic errors.

\section{DISCUSSION \label{discuss}}

\begin{figure}
\plotone{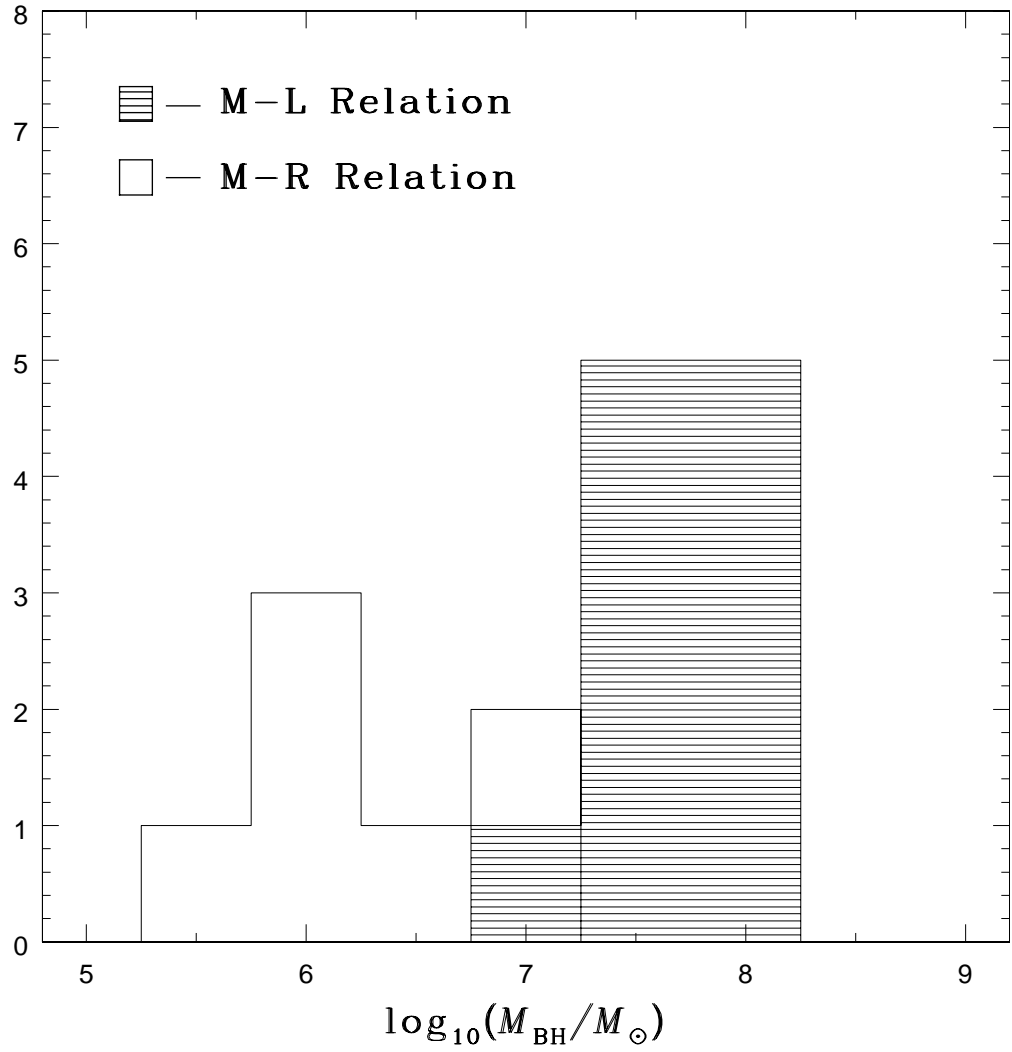} \figcaption{Histogram showing the
distribution of black hole masses derived in this study.
\label{histogram}}
\end{figure}

The black-hole masses presented in the previous section are plotted as
a histogram in Figure~\ref{histogram}.  There is a clear offset
between values obtained using different correlations.  We find that
the estimates of $\MBH$ calculated from the near-IR bulge luminosities
are, on average, more than one order of magnitude larger than those
obtained from the mass-radius relation.  The measurement uncertainty
for each object is typically 0.25 dex, which is insufficient to
account for the difference in the results.  Therefore, one (or more)
of the assumptions made in formulating the correlations must not be
applicable to this particular subset of active galaxies.  In the
following subsections, we consider the evidence supporting the
extension of each correlation to NLS1s, and discuss the impact of
these results on our understanding of these objects.

\subsection{Are Black Holes Less Massive In NLS1s? \label{undermass}}

The most common interpretation of the observed spectral
characteristics of NLS1s is that the central black holes are less
massive than those found in broad-line Seyfert galaxies.  This
hypothesis is supported in the literature by studies that employ the
mass-radius correlation as calibrated by reverberation mapping studies
\citep[e.g.][]{gru04a}.  Indeed, we find that for our sample the mean
value of $\MBH$ determined from (\ref{kaspi}) and (\ref{virial}) is
$\langle{\rm log}(\MBH)\rangle$ = 6.4.  If we accept the virial mass
estimates obtained from the mass-radius relation, the results from our
profile decompositions provide evidence that the galaxy bulges in our
sample are more luminous than predicted by the mass-luminosity
relation.  This is illustrated in Figures~\ref{bh_bulge_j} and
\ref{bh_bulge_k}, in which we compare the NLS1 black-hole masses as
determined from the mass-luminosity relation and the mass-radius
relation.  The estimates of $\MBH$ derived from the mass-radius
relation lie well below the best-fit line calculated from ellipticals
with secure $\MBH$ estimates.

It has been proposed that the departure of NLS1s from the
mass-luminosity relation is the result of the contribution of excess
flux from either bursts of nuclear star formation or AGN emission
\citep{nel04}.  We exercised great care in subtracting the nuclear
component from our high-resolution imaging data, however.  Any
residual emission inadvertently included in measuring the bulge
magnitude would be negligible, and certainly would not alter the
subsequent estimates of $\MBH$ by one full order of magnitude.
Rather, any errors in the nuclear subtraction routine would likely
involve an overcorrection, leading to underestimates of the black-hole
masses.  The estimated bulge scale radii are typically larger than
0.5\arcsec, and all are greater than the nominal PSF FWHM of
0.2\arcsec.  Therefore, we are confident that the derived parameters,
including the bulge magnitudes, are both physically reasonable and
accurate.

If black holes are less massive in NLS1s, there must be a physical
explanation for the breakdown of the mass-luminosity relation.  One
interpretation is that there is an inherent time delay between an
increase in the mass of the bulge through star formation and the
growth of the central black hole due to accretion.  In this scenario,
sufficient time passes after a gravitational interaction between
galaxies for the bulge to virialize and stars to form before gas
inflow triggers significant nuclear activity.  One could imagine that
accretion onto the central black hole is then halted by feedback
mechanisms as the mass becomes consistent with that of a ``normal''
Seyfert galaxy.  Numerical simulations of interactions by
\citet{dim05} suggest that such galactic nuclei undergo a period of
rapid accretion towards the end of a merger event due to the elevated
gas content near the core.  It is during this phase that the central
black holes can increase in mass by a significant fraction, and the
target emerges as an AGN.  During this period of rapid growth, the
accretion rate can approach the Eddington limit.  If NLS1s are indeed
evolutionarily young objects at the start of the AGN duty cycle, these
simulations constrain the NLS1 phase to a time relatively soon after a
gravitational interaction, a time when one might expect significant
morphological asymmetries in the host galaxies and particularly the
centers.

In order to search for such morphological evidence, we divided the
reduced {\it $K^{\prime}$}-band mosaics by the associated {\it J}-band
images.  We then applied an unsharp mask by convolving the quotient
image with a Gaussian function with a standard deviation of
1.5\arcsec.  No evidence was found of light asymmetries or tidal
features that would be indicative of a late-stage merger.
Furthermore, we note that spectroscopic detections of Seyfert nuclei
in systems currently undergoing a strong interaction
\citep[e.g.][]{liu95,van98} are difficult to explain if bulge growth
and AGN fueling are temporally well separated.

\begin{figure}
\plotone{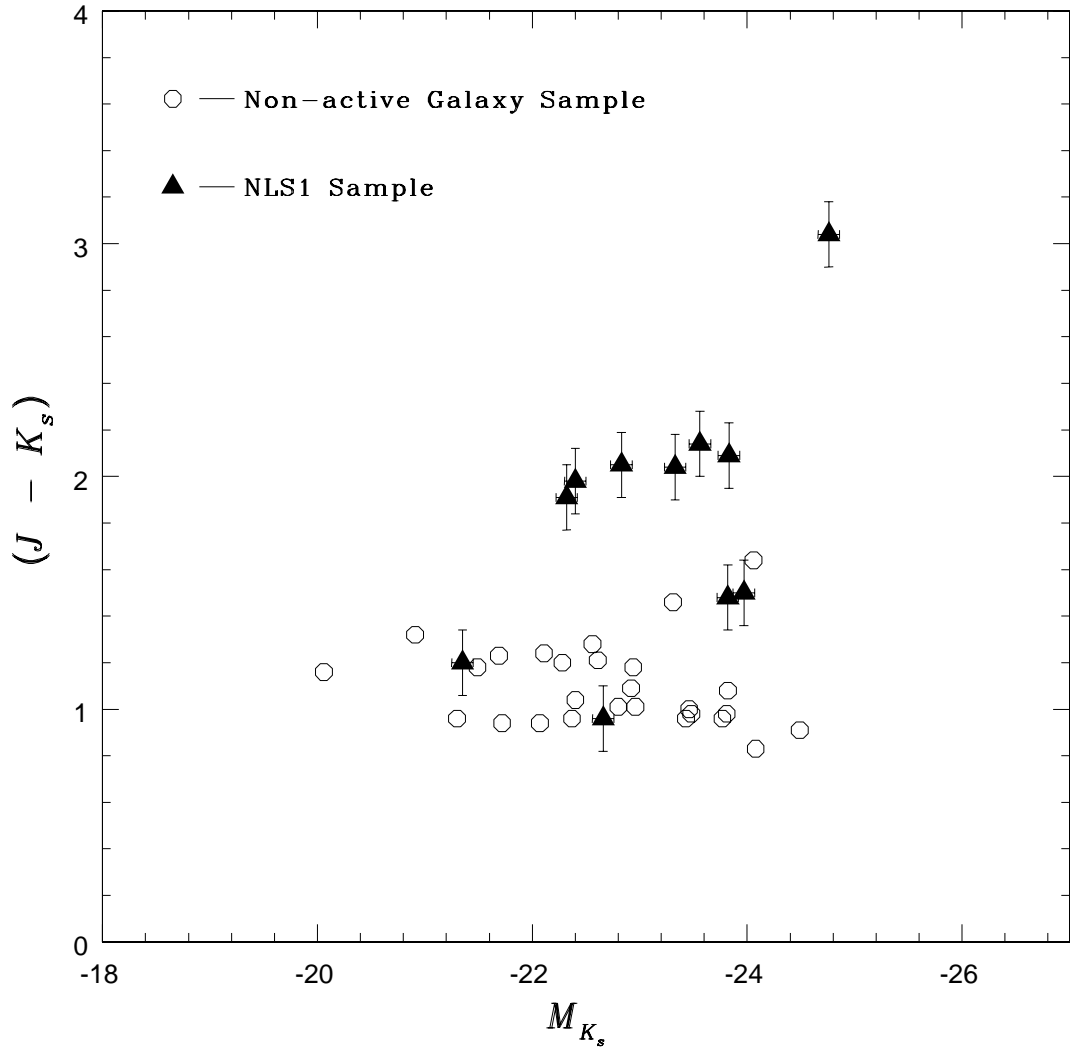} \figcaption{Comparison of the
bulge ({\it J}-{\it K$_s$}) colors for a sample of non-active galaxies
and NLS1s.  Open circles denote the non-active galaxy sample.  NLS1
galaxies are denoted by filled triangles.
\label{colors}}
\end{figure}

To further investigate the claim that NLS1s are evolutionarily young,
we compared the near-IR bulge colors of the objects in our study to a
subset of non-active, early-type galaxies taken from the sample
examined by \citet{don06}.  Three non-active galaxies were selected
for each NLS1, matched according to morphological class and absolute
magnitude of the total galaxy.  Of the 27 objects selected, we note
that nine are actually identified as containing a LINER nucleus.  The
mean bulge color for this subset of galaxies is equivalent to the mean
for the other galaxies in our non-active sample.  Therefore, their
inclusion does not adversely affect our results.  Using 2MASS XSC
imaging data, the ({\it J}-{\it K$_s$}) colors of the bulge components
of the non-active objects were computed with GALFIT.  The galaxy
parameters derived from the fits, including bulge magnitudes, are
presented in Table~\ref{inactiveset}.  The bulge ({\it J}-{\it K$_s$})
colors for the non-active galaxies and NLS1s are plotted against the
absolute bulge magnitudes in {\it K$_s$} in Figure~\ref{colors}.  The
mean bulge color for the non-active sample is ({\it J}-{\it K$_s$}) =
+1.10 $\pm$ 0.04, while for the NLS1 sample, the mean color was found
to be ({\it J}-{\it K$_s$}) = +1.85 $\pm$ 0.58.

In the absence of two-dimensional profile decompositions for a
matching sample of broad-line Seyferts, we compare our NLS1 colors to
results culled from the literature.  \citet{hun97} measured the
near-IR colors of galaxies hosting Seyfert and starburst nuclei.  The
mean color of the inner 3 kpc of the Seyfert 1 galaxies in their
sample was found to be ({\it J}-{\it K}) $\simeq$ +1.  However, the
inner 4\arcsec~were masked to exclude contamination from the active
nucleus; hence the measured colors more accurately represent the inner
disk component rather than the bulge.  Their use of an
early-generation InSb 62x58 IR array also limits the comparison to our
high-spatial resolution, AO-corrected data.

A comparison of the near-IR colors of Seyfert and non-active spiral
galaxies is presented in \citet{mar00}.  Using 1-dimensional profile
decompositions, the authors conclude that the bulge and disk
components of the two galaxy samples are similar in terms of their
mean ({\it J}-{\it K}) color and Fundamental Plane relation.  However
there is significant scatter in the measured bulge colors of the 18
AGN in their sample, and while the mean bulge color was determined to
be ({\it J}-{\it K}) = +1.25, several galaxies have a bulge color in
excess of +1.5.  One object in particular, NGC 4253, was found to have
an exceptionally red bulge, with ({\it J}-{\it K}) = +4.4.
Interestingly, while this galaxy is classified as a Seyfert 2 in
\citet{mar00}, it is identified as a NLS1 in \citet{cre03}.

We find that of the 11 NLS1s in our sample, four have a bulge color of
({\it J}-{\it K$_s$}) $\simeq$ +1 to +1.5, consistent with the sample
of non-active galaxies and other studies of broad-line Seyfert
galaxies.  However, six of the targets have a bulge color of ({\it
J}-{\it K$_s$}) $\simeq$ +1.9, and for one, Mrk 1239, we measure ({\it
J}-{\it K$_s$}) = +3.0.  We find no evidence for systematic effects
that would lead to errors in our measured colors; on the contrary, the
targets with the reddest bulge colors are also the model fits with the
lowest residuals, as parameterized by the reduced $\chi^2$ values.

The near-IR spectrum of Mrk 1239 may provide a clue to the source of
the unusual bulge colors.  A significant bump in nuclear continuum
emission, centered on 2.2 \micron, was detected by \citet{rod06}.
Given the resolution limit of approximately 1\arcsec~for their data,
the authors constrain the source of this emission to the central 380
pc, which is well outside the seeing disk of our imaging data.  A
similar flux excess has been detected in the NLS1s Mrk 766
\citep{rod05} and I Zw 1 \citep{rud00}.  Comparing the spectrum of Mrk
1239 to that of the NLS1 Ark 564, they find the latter does not
exhibit any excess continuum emission in the {\it K}-band.  Hence the
feature appears to be present in some, but curiously not all, NLS1
galaxies.  This may explain the large scatter in the ({\it J}-{\it
K$_s$}) bulge colors of our sample.

Because the observed {\it K$_s$}-band excess is constrained to the
central regions of the targets, we closely examined the inner bulge
colors of our targets.  We find that some of the NLS1s in our sample
exhibit a gradient, such that the bulge colors become redder with
decreasing radius.  This is consistent with the view that an
additional source of flux is detected at 2.2 \micron~in the inner
regions of some of our NLS1 targets.  Consequently, our {\it
K$_s$}-band bulge profiles may be influenced by this component,
producing the small but systematic residual pattern evident in
Figure~\ref{kprofile}.  Fitting the peak with a blackbody temperature
of 1200 K, \citet{rod06} suggest that the source of the excess flux is
hot graphite dust emission near the active nucleus.  This explanation,
however, cannot account for the spatial extent of the emission as
demonstrated by the observed color gradients.

We also note that the estimates of $\MBH$ derived from the profiles in
{\it K$_s$} are generally larger than those obtained from the {\it
J}-band fits.  Given the possible presence of an additional source of
flux near the centers of some NLS1s, we consider the mass estimates in
{\it J} to be more robust, and caution against the use of {\it
K$_s$}-band data to measure NLS1 host bulge luminosities.

In summary, while we cannot formally exclude the low-mass
interpretation, we find no evidence from our data that NLS1s as a
group have morphological asymmetries or extra stellar components
distinct from normal Seyfert galaxies that would suggest that these
objects contain a recently formed AGN.  It may well be that there is a
modest flux excess at 2.2 \micron~that affects the near-IR bulge
colors of some, but not all, NLS1 galaxies.

\subsection{Could The Mass-Luminosity Relation Hold For NLS1s? \label{masslum2}}

The alternative interpretation of our results is that the
mass-luminosity relation does indeed hold for NLS1s and that the
masses obtained from the correlation with broad-line radius are
unreasonably low.  It is evident from Figures~\ref{bh_bulge_j} and
\ref{bh_bulge_k} that the measured NLS1 bulge magnitudes are similar
to those of the non-active galaxies used to derive the correlations.
The estimates of $\MBH$ for the NLS1s were obtained by interpolating
within a region for which the correlations are known to be valid for
broad-line Seyferts.  This result is not in agreement with
\citet{bot04}, who suggest that the {\it B}-band luminosities of NLS1
bulges are lower than those of the broad line counterparts by
approximately 1.3 magnitudes.  However, as discussed earlier, near-IR
wavelengths should in principle provide a more accurate trace of the
underlying mass distribution than the {\it B}-band.

We are confident that our spectroscopic measurements do not suffer
from any systematic errors.  Therefore, if we accept the extension of
the mass-luminosity relation to NLS1s, we require an explanation for
the narrow emission lines that characterize these objects.  Although
strong non-virial motions in the broad-line region are known to induce
asymmetries in the $\C$ line profiles in AGN spectra, their
contribution at optical wavelengths is insufficient to significantly
affect the H$\beta$ line profile \citep{ver01b}.  We have yet to
directly resolve the broad line region, and rely on indirect evidence
that the motion of the gas is virialized \citep{fer05}.  If nuclear
activity is triggered by galaxy mergers, the gas may be heavily
influenced by non-gravitational forces, particularly at the earliest
stages of the AGN duty cycle \citep[e.g.][]{bar91}.  It is possible
that the broad-line emitting gas in the cores of NLS1s does not
reflect only the gravitational potential of the central source.
Departures from virialized motions would almost certainly broaden
emission lines, increasing the discrepancy with the mass-luminosity
relation.

As noted by \citet{pet00b}, there are potential explanations for the
remarkable spectral properties of NLS1s other than the low-mass
hypothesis.  It is possible that the orientation of the broad-line
region could influence mass estimates derived using the \citet{kas05}
relation.  Studies of the central engines of active galaxies typically
involve a large sample of objects with a range of spectral properties,
and the assumption that the broad-line regions are oriented randomly
does not affect the statistical results.  On the other hand, NLS1s are
a small and distinct subset of Seyfert galaxies, and the assumption of
heterogeneity may not be reasonable.  If NLS1 systems preferentially
have a low inclination, values of $\MBH$ could be underestimated by a
factor of three or more.

In \S~\ref{massrad}, it was mentioned that the model geometry of the
broad-line region could influence estimates of $\MBH$.  An isotropic
gas distribution is assumed in many studies that employ the
mass-radius relation \citep[e.g.][]{kas00,wan02}.  The exclusion of a
disk component in these studies could produce virial black-hole masses
that are a underestimated by a factor of three.  As the two effects
mentioned here are multiplicative, their net effect could lead to
values of $\MBH$ that are a full order of magnitude too low.
Therefore, it is possible that our incongruent mass estimates are the
result of the combined effects of broad-line geometry and orientation,
and the mass-luminosity relation could indeed hold for NLS1s.

\subsection{Are NLS1s Fueled By Merger Events? \label{secular}}

Given the discrepancy between the estimates of $\MBH$ revealed in this
study, it is apparent that we cannot apply both the mass-radius and
mass-luminosity relations to the NLS1 subset in the same form as used
for broad-line AGN and non-active galaxies.  The most common
interpretation is a breakdown in the correlation with bulge luminosity.
We have exercised a great deal of care in measuring accurate bulge
magnitudes.  We also have demonstrated that the application of the
mass-radius relation to NLS1s may suffer from systematic effects that
could lead to underestimated virial masses.  Furthermore, we find no
evidence for features that betray recent tidal interactions or
distinct stellar components that would support the interpretation that
NLS1s are evolutionarily young.

There may be an alternative view of NLS1s that provides an explanation
for their exceptional properties, however.  There are some
interesting, albeit circumstantial, lines of evidence that point to a
mechanism other than galaxy harassment to initiate accretion in these
objects.  It has been suggested that as the universe continues to
expand and cosmic distances increase the role of mergers in defining
galaxy morphology will become less important compared to secular
evolution, the gradual reorganization of mass due to internal
processes such as bar-driven instabilities \citep{kor04}.  The
photometric and kinematic properties of galaxies that have undergone
secular evolution will be distinct and recognizable from those which
arise from merger events.  It is conceivable that NLS1s represent such
objects.

The first piece of evidence in favor of this hypothesis involves the
shape of the bulge surface brightness profiles.  Bulges formed via
secular evolution are believed to be more accurately represented by
exponential profiles than by the de Vaucouleurs $r^{1/4}$ law
\citep{bal03}.  When fitting the bulge components for the galaxies in
our sample, the S\'{e}rsic index, {\it n}, was allowed to vary as a
free parameter within the GALFIT minimization routine.  The average of
the optimal indices determined by GALFIT was 1.4 in {\it $K^{\prime}$}
and 1.6 in {\it J}, and {\it n} was never larger than 2.4.  The NLS1
bulges were thus found to be more consistent with an exponential
profile.

To be fair, the true nature of bulges remains somewhat ambiguous.  In
studying the bulge profiles of S0 to Sbc galaxies using the NICMOS
near-IR camera on the Hubble Space Telescope, \citet{bal03} found the
average S\'{e}rsic index to be 1.7 $\pm$ 0.7.  Ground-based near-IR
observations of the same sample were best fit by indices of 4 to 6.
The authors suggest that the disagreement is caused by nuclear point
source blending due to the effects of atmospheric seeing in the
ground-based data.  It is possible that this problem was mitigated in
our ground-based data by the excellent angular resolution, combined
with the application of our empirical technique for removing the
nuclear emission.
	
Another key piece of evidence supporting secular evolution is the
possible failure of the $\MBH$-$\sigma_s$ correlation in NLS1s.
Numerical simulations by \citet{dim05} demonstrate that major galaxy
mergers produce strong inflows of gas that feed the black hole.  The
energy emitted by the black hole expels gas from the central region,
limiting star formation and regulating further accretion.  The
connection between $\MBH$ and bulge velocity dispersion could
therefore be a natural product of galaxy mergers.  Conversely, the
departure from this relationship by NLS1s may indicate that mergers
are not largely responsible for initiating nuclear activity.  Geometry
may thus play a more significant role in the narrow H$\beta$ emission
lines.

If accretion in NLS1s is not primarily the result of gravitational
interactions, there must be an alternative mechanism for driving gas
towards the nucleus.  In the secular evolution paradigm, this is
accomplished by stellar bars that produce non-axisymmetric
gravitational potentials, driving gas inwards.  Thus, barred
early-type galaxies are ideal candidates for secular evolution
\citep{kor04}.  Interestingly, it was reported by \citet{cre03} that
large-scale bars appear to be more common in NLS1s than in broad-line
AGN.  Of the seven objects in our sample for which morphological
information was available, four are classified as barred galaxies.  We
did not detect any such features in our near-IR data that could be
included in the decompositions, however.  Observations of early-type
barred and unbarred systems by \citet{agu05} demonstrate that the
presence of a bar does not change the galaxy morphological parameters
such as S\'{e}rsic index and effective radius.

Given the arguments presented above, we postulate that NLS1s may
represent AGN in which accretion may be regulated by internal
processes rather than galaxy mergers.  While there are outstanding
issues that have not been addressed in this hypothesis, such as the
x-ray properties of NLS1s and the exact cause of the narrow emission
features, we feel that further investigation into the possibility of
secular development is warranted.

\section{CONCLUSIONS \label{concl}}

We have studied a sample of NLS1 galaxies in order to explore the
hypothesis that such objects are evolutionarily young AGN possessing
relatively low-mass black holes.  Galaxy profiles were separated into
their constituent components by applying the two-dimensional
decomposition algorithm GALFIT to high-spatial resolution, near-IR
imaging data.  Black-hole masses were derived using the correlation
between $\MBH$ and host bulge luminosity, calibrated in {\it J} and
{\it $K^{\prime}$} using a sample of bright elliptical galaxies with
black-hole masses obtained by resolving the sphere of influence.  The
scatter in the mass-luminosity relation was thus minimized by
restricting the calibration sample to objects with secure $\MBH$
estimates.  Furthermore, the near-IR form of the relation is less
sensitive to attenuation by dust than at optical wavelengths.  The
unweighted mean black-hole mass determined from this relation is, in
solar units, $\langle{\rm log}(\MBH)\rangle$ = 7.9, in line with
typical values for broad-line Seyferts.  We also derived masses for
seven of the objects in our sample from the correlation between the
radius of the broad-line region and the monochromatic continuum
luminosity at 5100 \AA~under the assumption that the emission-line gas
is virialized and using optical spectra taken from the literature.
The mean mass calculated from this relation is $\langle{\rm
log}(\MBH)\rangle$ = 6.4, more consistent with the premise that the
black holes in NLS1s are less massive than those found in their
broad-line counterparts.

It is clear that there is a significant discrepancy between the
results obtained from these correlations, with the mass-luminosity
relation yielding values of $\MBH$ that are on average more than one
order of magnitude larger than those obtained from the mass-radius
relation.  Formal uncertainties in the calculated masses are typically
around 0.25 dex, and thus are insufficient to account for the
conflicting results.  Our findings suggest that (at least) one of the
correlations cannot be extended to the NLS1 subset in the same form as
used for other galaxies.  It has been conjectured that NLS1s may not
follow the standard mass-luminosity relation due to contamination by
circumnuclear star formation.  We have argued that this extra
component would not significantly affect our data.  Furthermore, we
noted that the resolution of our imaging data provided excellent
spatial over-sampling of the surface brightness profiles, producing
highly accurate bulge luminosities.  The resulting $\MBH$ estimates
were calculated by interpolating the relation within a region known to
be valid for both non-active galaxies and other AGN.

Numerical simulations of galaxy interactions constrain the
high-accretion rate period that may define the start of the AGN duty
cycle to a period shortly after a significant gravitational
perturbation of the host galaxy.  To further investigate this
possibility, we compared the NLS1 bulge ({\it J}-{\it K$_s$}) color
indices to a matched sample of non-active galaxies.  We determined the
mean bulge color for the NLS1 sample to be $\langle({\it J}-{\it
K_s})\rangle$ = +1.85 $\pm$ 0.58.  For the non-active galaxy sample,
the mean color is $\langle({\it J}-{\it K_s})\rangle$ = +1.10 $\pm$
0.04.  Estimates of the bulge colors for broad-line Seyfert 1 galaxies
were taken from the literature.  A fair comparison to our sample is
complicated by the large scatter in the published results and the
potential for misidentification of NLS1s as Seyfert 2 galaxies.  The
NLS1 images were also examined for light asymmetries or tidal features
that may reflect a recent gravitational interaction.  We find no
evidence for any significant morphological asymmetries or extra
stellar components that would suggest these galaxies are
evolutionarily young.

We have, however, identified several lines of circumstantial evidence
suggesting that nuclear fueling in NLS1s may be strongly influenced by
secular processes within the host galaxy.  The bulge light profiles
obtained from our two-dimensional decompositions are all nearly
exponential in shape, a key characteristic of secular evolution.
NLS1s may not follow the standard $\MBH$-$\sigma_s$ relation, which
has been shown to be a natural product of galaxy mergers, while an
enhanced fraction of stellar bar structures in NLS1s may provide a
natural source for fueling the nucleus without invoking a recent
gravitational interaction.  It is thus fair to acknowledge that we do
not yet completely understand the evolution of this important subset
of low-redshift active galaxies.

\acknowledgements

ACKNOWLEDGMENTS.

The authors would like to thank the anonymous referee for insightful
comments that improved our manuscript.  The authors would also like to
thank the Natural Sciences and Engineering Research Council of Canada
and the CFHT TAC for supporting this work, and X.Y. Dong for
insightful discussions.  This publication makes use of data products
from the Two Micron All Sky Survey, which is a joint project of the
University of Massachusetts and the Infrared Processing and Analysis
Center/California Institute of Technology, funded by the National
Aeronautics and Space Administration and the National Science
Foundation.  This research has made use of the NASA/IPAC Extragalactic
Database (NED) which is operated by the Jet Propulsion Laboratory,
California Institute of Technology, under contract with the National
Aeronautics and Space Administration.

%\clearpage

% ** Tables here **

\begin{deluxetable}{l c c c c c c}
%\tabletypesize{\small}
\tabletypesize{\scriptsize}
%\rotate
\tablecaption{NLS1 Galaxy Parameters\label{params}}
\tablewidth{0pt}
\tablehead{
  \colhead{Target ID} &
  \colhead{Distance\tablenotemark{a}} &
  \colhead{Morph. Class.} &
  \multicolumn{2}{c}{Total Magnitude\tablenotemark{b}} &
  \colhead{Optical SMA} & 
  \colhead{$\it{K}$=+20 Isophotal Radius} \\
  \colhead{} &
  \colhead{(Mpc)} &
  \colhead{} &
  \colhead{\it{J}} &
  \colhead{\it{K$_s$}} &
  \colhead{(arcsec)} & 
  \colhead{(arcsec)}
}
\startdata
Mrk 335 & 111.1 & S0/a & +12.03 $\pm$ 0.04 & +10.06 $\pm$ 0.03 & 9 & 15.1 \\
Mrk 359 & 74.8 & SB0a & +11.61 $\pm$ 0.02 & +10.46 $\pm$ 0.03 & 18 & 16.6 \\
Mrk 618 & 153.6 & SB(s)b pec & +11.82 $\pm$ 0.03 & +10.37 $\pm$ 0.03 & 27 & 18.1 \\
Mrk 705 & 125.7 & S0? & +11.94 $\pm$ 0.03 & +10.74 $\pm$ 0.04 & 21 & 16.1 \\
Mrk 734 & 217.6 & Compact & +13.04 $\pm$ 0.04 & +11.76 $\pm$ 0.06 & 18 & 7.6 \\
Mrk 1044 & 70.7 & SB0 & +11.78 $\pm$ 0.03 & +10.47 $\pm$ 0.03 & 21 & 14.0 \\
Mrk 1126 & 45.7 & (R)SB(r)a & +11.09 $\pm$ 0.02 & +10.14 $\pm$ 0.04 & 54 & 26.1 \\
Mrk 1239 & 85.8 & E-S0 & +11.87 $\pm$ 0.02 & +9.60 $\pm$ 0.02 & 12 & 10.1 \\
IRAS 04596-2257 & 176.5 & - & +13.21 $\pm$ 0.04 & +12.03 $\pm$ 0.06 & 9.3 & 6.4 \\
MCG 08.15.009 & 105.0 & - & +12.42 $\pm$ 0.05 & +11.28 $\pm$ 0.07 & 18 & 14.1 \\
PG 1016+336 & 105.6 & Compact & +13.03 $\pm$ 0.05 & +11.67 $\pm$ 0.06 & 9 & 11.6 \\
\enddata
\tablenotetext{a}{Distances were calculated using redshift information obtained from the literature.}

\tablenotetext{b}{Total galaxy magnitudes and isophotal radii at {\it
K} = +20 are taken from the 2MASS XSC; optical radii are taken from the literature.}
\end{deluxetable}

\begin{deluxetable}{l l l c c c c}
%\tabletypesize{\small}
\tabletypesize{\scriptsize}
%\rotate
\tablecaption{Summary of Observations\label{observations}}
\tablewidth{0pt}
\tablehead{
  \colhead{Target ID} &
  \multicolumn{2}{c}{Coordinates} &
  \multicolumn{2}{c}{Exposure Time\tablenotemark{a}} &
  \multicolumn{2}{c}{Mean Airmass} \\
  \colhead{} &
  \colhead{R.A.\,(J2000)} &
  \colhead{Decl.\,(J2000)} &
  \colhead{\it{J}} &
  \colhead{\it{K$^\prime$}} &
  \colhead{\it{J}} &
  \colhead{\it{K$^\prime$}}
}
\startdata
Mrk 335 & 00 06 19.5 & +20 12 10.5 & 1200 & 80 & 1.18 & 1.08 \\
Mrk 359 & 01 27 32.6 & +19 10 43.8 & 1200 & 480 & 1.04 & 1.13 \\
Mrk 618 & 04 36 22.2 & $-$10 22 33.8 & 1200 & 120 & 1.17 & 1.16 \\
Mrk 705 & 09 26 03.3 & +12 44 03.6 & 1680 & 480 & 1.04 & 1.14 \\
Mrk 734 & 11 21 47.1 & +11 44 18.3 & 480 & 240 & 1.01 & 1.02 \\
Mrk 1044 & 02 30 05.4 & $-$08 59 52.6 & 1920 & 480 & 1.16 & 1.27 \\
Mrk 1126 & 23 00 47.8 & $-$12 55 06.7 & 1680 & 1680 & 1.40 & 1.63 \\
Mrk 1239 & 09 52 19.1 & $-$01 36 43.5 & 480 & 240 & 1.09 & 1.09 \\
IRAS 04596-2257 & 05 01 48.6 & $-$22 53 23.2 & 1680 & 1920 & 1.41 & 1.44 \\
MCG 08.15.009 & 07 51 51.9 & +49 48 51.6 & 1920 & 1920 & 1.16 & 1.16 \\
PG 1016+336 & 10 19 49.5 & +33 22 03.7 & 1680 & 480 & 1.03 & 1.04 \\
\enddata
\tablecomments{Units of right ascension are hours, minutes, and
seconds; units of declination are degrees, arcminutes, and
arcseconds.}
\tablenotetext{a}{Exposure times represent total on-target integration time.}
\end{deluxetable}

\begin{deluxetable}{l c c c c c c c c c c}
%\tabletypesize{\small}
\tabletypesize{\scriptsize}
%\rotate
\tablecaption{Photometric and Structural Parameters of Components\label{galfitparams}}
\tablewidth{0pt}
\tablehead{
  \colhead{Target ID} &
  \multicolumn{5}{c}{$\it{J}$-band Results} &
  \multicolumn{5}{c}{$\it{K_s}$-band Results} \\
  \colhead{} &
  \colhead{{\it M}(bulge)\tablenotemark{a}} &
  \colhead{{\it M}(disk)} &
  \colhead{B/D} &
  \colhead{S\'{e}rsic $\it{n}$} &
  \colhead{r (kpc)\tablenotemark{b}} &
  \colhead{{\it M}(bulge)} &
  \colhead{{\it M}(disk)} &
  \colhead{B/D} &
  \colhead{S\'{e}rsic $\it{n}$} &
  \colhead{r (kpc)}
}
\startdata
Mrk 335$^\dag$ & -21.29 & -21.69 & 0.70 & 1.02 & 0.25 & -23.33 & -22.91 & 1.43 & 1.23 & 0.15 \\
Mrk 359 & -20.42 & -22.38 & 0.16 & 1.92 & 0.23 & -22.40 & -23.97 & 0.24 & 2.13 & 0.23 \\
Mrk 618 & -21.74 & -23.01 & 0.31 & 1.20 & 0.22 & -23.83 & -27.00 & 0.05 & 1.55 & 0.26 \\
Mrk 705$^\dag$ & -22.47 & -24.66 & 0.13 & 2.40 & 0.93 & -23.97 & -25.87 & 0.17 & 2.28 & 0.54 \\
Mrk 734$^\dag$ & -22.34 & -25.58 & 0.05 & 2.10 & 0.60 & -23.82 & -24.32 & 0.63 & 1.47 & 0.33 \\
Mrk 1044 & -20.78 & -21.35 & 0.59 & 0.96 & 0.23 & -22.83 & -22.49 & 1.36 & 0.83 & 0.17 \\
Mrk 1126 & -20.15 & -22.59 & 0.11 & 1.64 & 0.29 & -21.35 & -22.72 & 0.29 & 1.75 & 0.21 \\
Mrk 1239 & -21.72 & -22.63 & 0.43 & 1.65 & 0.17 & -24.76 & -25.60 & 0.46 & 1.08 & 0.19 \\
IRAS 04596-2257 & -21.42 & -21.90 & 0.64 & 1.16 & 0.37 & -23.56 & -23.49 & 1.08 & 0.81 & 0.60 \\
MCG 08.15.009 & -21.70 & - & - & 1.45 & 1.02 & -22.66 & - & - & 1.15 & 0.56 \\
PG 1016+336 & -20.41 & -21.06 & 0.55 & 1.27 & 0.19 & -22.32 & -22.05 & 1.29 & 0.94 & 0.12 \\
\enddata
\tablecomments{Galaxies with multiple image sets acquired in $\it{K_s}$ are denoted by $^\dag$.
For these objects, the mean fit parameters are listed.}
\tablenotetext{a}{Absolute magnitudes of galaxy components derived from GALFIT output.}
\tablenotetext{b}{Bulge scale radius in kpc.}
\end{deluxetable}
\clearpage

\begin{deluxetable}{l c c c c}
%\tabletypesize{\small}
\tabletypesize{\scriptsize}
\tablecaption{Galaxies Used to Calibrate Mass-Luminosity Relation.\label{ellipticals}}
\tablewidth{0pt}
\tablehead{
  \colhead{Object} &
  \colhead{Distance\tablenotemark{a}} &
  \colhead{${M_{BH}}$} &
  \colhead{$M_{J}$} &
  \colhead{$M_{K_s}$} \\
  \colhead{} &
  \colhead{(Mpc)} &
  \colhead{$(10^8\,M_{\sun})$} &
  \colhead{} &
  \colhead{}
  }
\startdata
IC 1459 & 28.8 & 26.0 $\pm$ 11.0 & -24.62 $\pm$ 0.28  & -25.50 $\pm$ 0.28 \\
NGC 221 & 0.77 & 0.025 $\pm$ 0.005 & -18.21 $\pm$ 0.08 & -19.36 $\pm$ 0.08 \\
NGC 3377 & 10.9 & $1.00^{+0.9}_{-0.1}$ & -21.93 $\pm$ 0.09 & -22.76 $\pm$ 0.09 \\
NGC 3379 & 10.3 & 1.35 $\pm$ 0.73 & -22.93 $\pm$ 0.11 & -23.81 $\pm$ 0.11 \\
NGC 3608 & 22.5 & $1.9^{+1.0}_{-0.6}$ & -22.83 $\pm$ 0.14 & -23.67 $\pm$ 0.14 \\
NGC 4261 & 31.1 & 5.4 $\pm$ 1.2 & -24.28 $\pm$ 0.19 & -25.21 $\pm$ 0.19 \\
NGC 4291 & 25.3 & $3.1^{+0.8}_{-2.3}$ & -22.69 $\pm$ 0.32 & -23.62 $\pm$ 0.32 \\
NGC 4374 & 17.7 & $17^{+12}_{-6.7}$ & -24.15 $\pm$ 0.11 & -25.04 $\pm$ 0.11 \\
NGC 4473 & 15.3 & $1.1^{+0.5}_{-0.8}$ & -22.91 $\pm$ 0.13 & -23.78 $\pm$ 0.13 \\
NGC 4486 & 15.8 & 35.7 $\pm$ 10.2 & -24.29 $\pm$ 0.16 & -25.18 $\pm$ 0.16 \\
NGC 4564 & 14.5 & $0.56^{+0.03}_{-0.08}$ & -21.97 $\pm$ 0.17 & -22.89 $\pm$ 0.17 \\
NGC 4649 & 16.4 & $20.0^{+4.0}_{-6.0}$ & -24.43 $\pm$ 0.15 & -25.35 $\pm$ 0.15 \\
NGC 4697 & 11.4 & $1.7^{+0.2}_{-0.3}$ & -23.08 $\pm$ 0.14 & -23.94 $\pm$ 0.14 \\
NGC 5845 & 24.7 & $2.4^{+0.4}_{-1.4}$ & -21.94 $\pm$ 0.21 & -22.88 $\pm$ 0.21 \\
NGC 6251$^\dag$ & 99.4 & 5.9 $\pm$ 2.0 & -25.01 $\pm$ 0.22 & -25.99 $\pm$ 0.22 \\
NGC 7052$^\dag$ & 62.5 & $3.7^{+2.6}_{-1.5}$ & -24.55 $\pm$ 0.22 & -25.45 $\pm$ 0.22 \\
\enddata 

\tablenotetext{a}{Galaxy distances obtained from \citet{ton01} except
for two objects denoted by $^\dag$.  Distances for these galaxies
determined using redshift information from the RC3.}
\end{deluxetable}
\clearpage

\begin{deluxetable}{l c c c c c c c c}
%\rotate
%\tabletypesize{\small}
\tabletypesize{\scriptsize}
\tablecaption{Black Hole Mass Estimates\label{bhmasses}}
\tablewidth{0pt}
\tablehead{
  \colhead{Target ID} &
  \colhead{FWHM(H$\beta$)\tablenotemark{a}} &
  \colhead{$\it \lambda L_{\lambda}$(5100~{\rm\AA})} &
  \colhead{{\it L}(H$\beta$)} &
  \multicolumn{2}{c}{log $\MBH$(Mass-Radius Relation)\tablenotemark{b}} &
  \multicolumn{3}{c}{log $\MBH$(Mass-Luminosity Relation)} \\
  \colhead{} &
  \colhead{(km s$^{-1}$)} &
  \colhead{(10$^{44}$ erg s$^{-1}$)} &
  \colhead{(10$^{42}$ erg s$^{-1}$)} &
  \colhead{$\it L_{\lambda}$(5100~{\rm\AA})} &
  \colhead{{\it L}(H$\beta$)} &
  \colhead{$\it{J}$-band} &
  \colhead{$\it{K_s}$-band} &
  \colhead{T Type}
}

\startdata
Mrk 335 & 1640 & 0.640 $\pm$ 0.006 & 1.0 $\pm$ 0.1 & 6.81 & 6.99 & 7.64 & 8.12 & 8.37 \\
Mrk 359 & 480 & 0.193 $\pm$ 0.004 & 0.07 $\pm$ 0.01 & 5.38 & 5.28 & 7.27 & 7.71 & 7.82 \\
Mrk 618 & 2300 & - & - & - & - & 7.83 & 8.34 & 8.31 \\
Mrk 705 & 1990 & 0.273 $\pm$ 0.004 & 0.33 $\pm$ 0.04 & 6.72 & 6.88 & 8.15 & 8.40 & 8.26 \\
Mrk 734 & 1820 & 0.728 $\pm$ 0.007 & 0.6 $\pm$ 0.1 & 6.94 & 6.94 & 8.09 & 8.33 & - \\
Mrk 1044 & 1280 & 0.153 $\pm$ 0.003 & 0.19 $\pm$ 0.02 & 6.17 & 6.36 & 7.42 & 7.90 & 7.76 \\
Mrk 1126 & 2500 & - & - & - & - & 7.15 & 7.25 & 7.49 \\
Mrk 1239 & 910 & 0.163 $\pm$ 0.004 & 0.24 $\pm$ 0.04 & 5.89 & 6.13 & 7.83 & 8.75 & 8.41 \\
IRAS 04596-2257 & 1500 & - & - & - & - & 7.70 & 8.22 & - \\
MCG 08.15.009 & - & - & - & - & - & 7.82 & 7.82 & 7.92 \\
PG 1016+336 & 1600 & 0.073 $\pm$ 0.001 & 0.059 $\pm$ 0.006 & 6.14 & 6.28 & 7.26 & 7.68 & 7.61 \\
\tableline
Average & & & & 6.29 & 6.41 & 7.65 & 8.05 & 7.99 \\
\enddata 
\tablenotetext{a}{Values for the FWHM of the broad component
of H$\beta$ taken from \citet{ver01b}.}
\tablenotetext{b}{Black-hole masses are presented in log space, in
units of solar masses.}
\end{deluxetable}
\clearpage

\begin{deluxetable}{l c c c c c}
%\tabletypesize{\small}
\tabletypesize{\scriptsize}
\tablecaption{Non-active Galaxy Parameters\label{inactiveset}}
\tablewidth{0pt}
\tablehead{
  \colhead{Target ID} &
  \colhead{Distance\tablenotemark{a}} &
  \multicolumn{2}{c}{$\it{J}$-band Results} &
  \multicolumn{2}{c}{$\it{K_s}$-band Results} \\
  \colhead{} &
  \colhead{(Mpc)} &
  \colhead{{\it M}(bulge)\tablenotemark{b}} &
  \colhead{{\it M}(disk)} &
  \colhead{{\it M}(bulge)} &
  \colhead{{\it M}(disk)}
}
\startdata
IC 356 & 18.1 & -22.83 & -24.29 & -23.81 & -25.12 \\
IC 520 & 47.0 & -22.50 & -23.32 & -23.48 & -24.22 \\
NGC 266 & 62.4 & -21.85 & -24.08 & -23.31 & -24.88 \\
NGC 2146 & 17.2 & -22.42 & -22.10 & -24.06 & -21.47 \\
NGC 2775 & 17.0 & -21.08 & -23.10 & -22.28 & -23.91 \\
NGC 2782 & 37.3 & -21.28 & -22.61 & -22.56 & -23.32 \\
NGC 2985 & 22.4 & -22.81 & -22.59 & -23.77 & -23.41 \\
NGC 3166 & 22.0 & -21.83 & -22.64 & -22.92 & -23.40 \\
NGC 3169 & 19.7 & -22.74 & -22.34 & -23.82 & -22.93 \\
NGC 3190 & 22.4 & -21.76 & -22.41 & -22.94 & -23.47 \\
NGC 3504 & 26.5 & -21.40 & -22.60 & -22.61 & -23.44 \\
NGC 3583 & 34.0 & -20.46 & -22.33 & -21.69 & -23.42 \\
NGC 3705 & 17.0 & -21.36 & -21.11 & -22.40 & -21.88 \\
NGC 3729 & 17.0 & -18.90 & -20.79 & -20.06 & -21.65 \\
NGC 3884 & 91.6 & -22.46 & -24.20 & -23.46 & -25.13 \\
NGC 3898 & 21.9 & -22.47 & -22.28 & -23.43 & -23.10 \\
NGC 4064 & 16.9 & -20.31 & -21.38 & -21.49 & -22.05 \\
NGC 4192 & 16.8 & -20.87 & -23.21 & -22.11 & -24.07 \\
NGC 4293 & 17.0 & -19.59 & -21.73 & -20.91 & -22.76 \\
NGC 4369 & 21.6 & -20.34 & -21.63 & -21.30 & -22.44 \\
NGC 4457 & 17.4 & -21.79 & -21.97 & -22.80 & -22.68 \\
NGC 4643 & 25.7 & -23.58 & -22.11 & -24.49 & -23.07 \\
NGC 4665 & 17.9 & -21.13 & -22.67 & -22.07 & -23.44 \\
NGC 5377 & 31.0 & -21.95 & -22.85 & -22.96 & -23.68 \\
NGC 5701 & 26.1 & -23.25 & -20.66 & -24.08 & -21.97 \\
NGC 6340 & 22.0 & -20.78 & -22.21 & -21.72 & -23.03 \\
NGC 6654 & 29.5 & -21.41 & -22.60 & -22.37 & -23.47 \\
\enddata

\tablenotetext{a}{Distances were calculated using redshift information
obtained from the literature.}
\tablenotetext{b}{Absolute magnitudes of components derived from
GALFIT model fits to 2MASS data.}
\end{deluxetable}
\clearpage

%Figures

\begin{figure}
\plotone{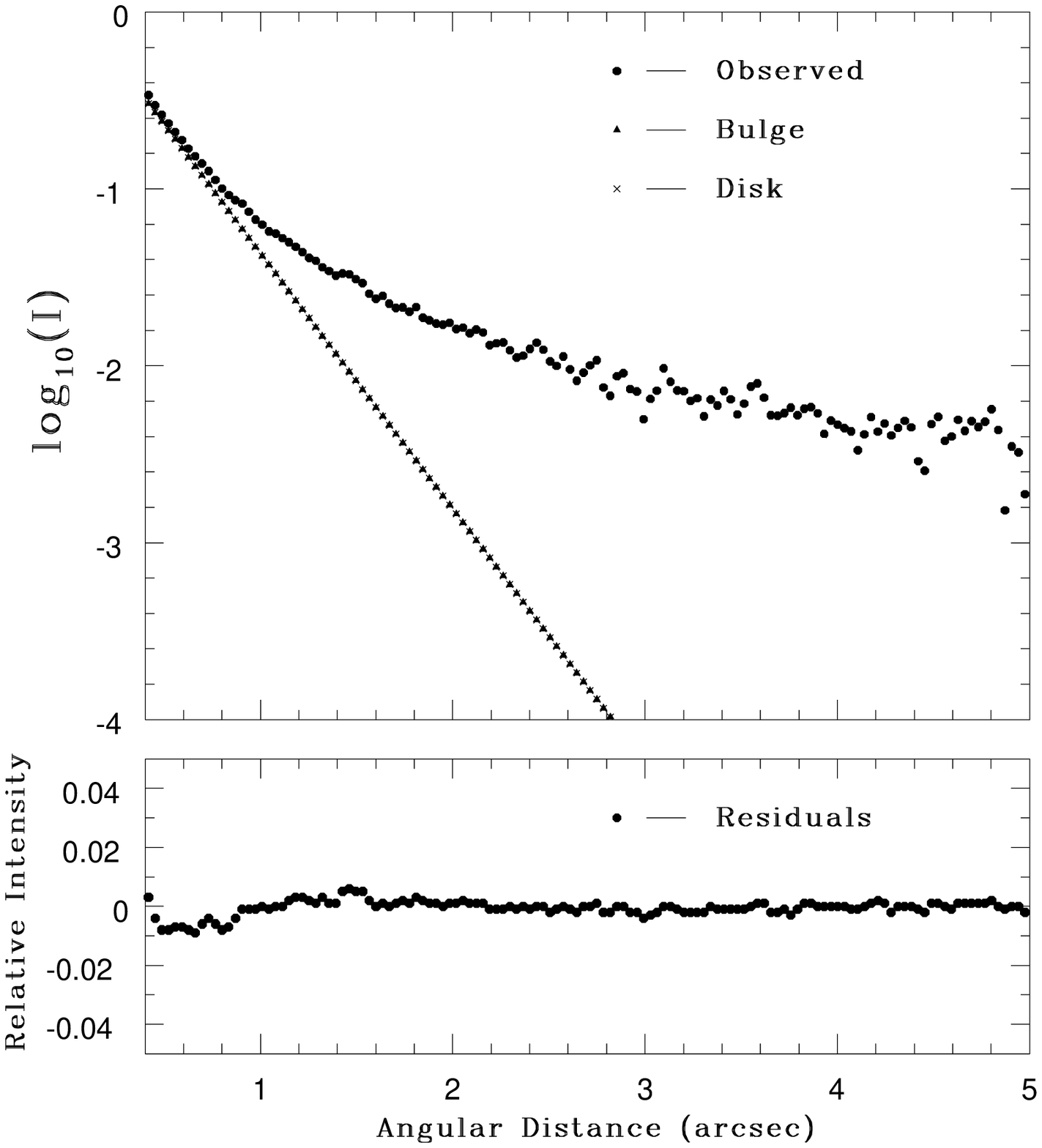} \figcaption{{\it J}-band surface brightness profile
for Mrk 335.}
\end{figure}
\begin{figure}
\plotone{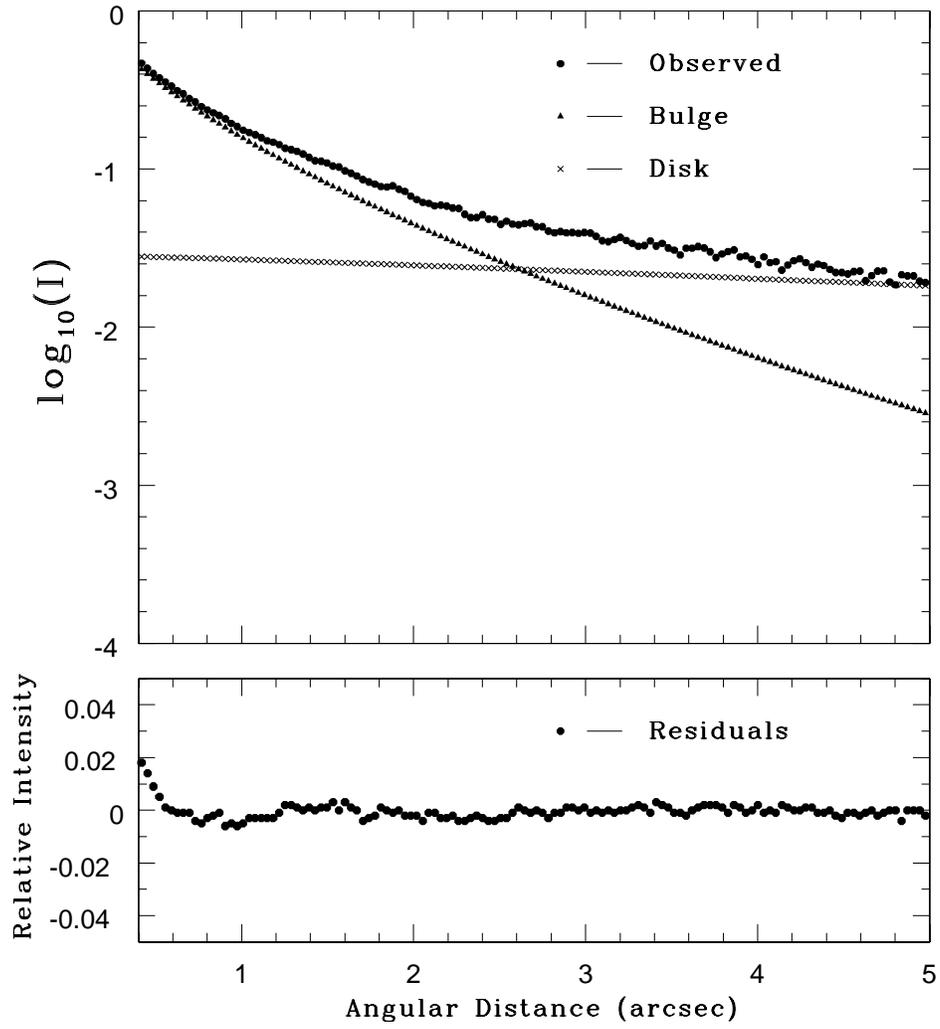} \figcaption{{\it J}-band surface brightness profile
for Mrk 1126.}
\end{figure}
\begin{figure}
\plotone{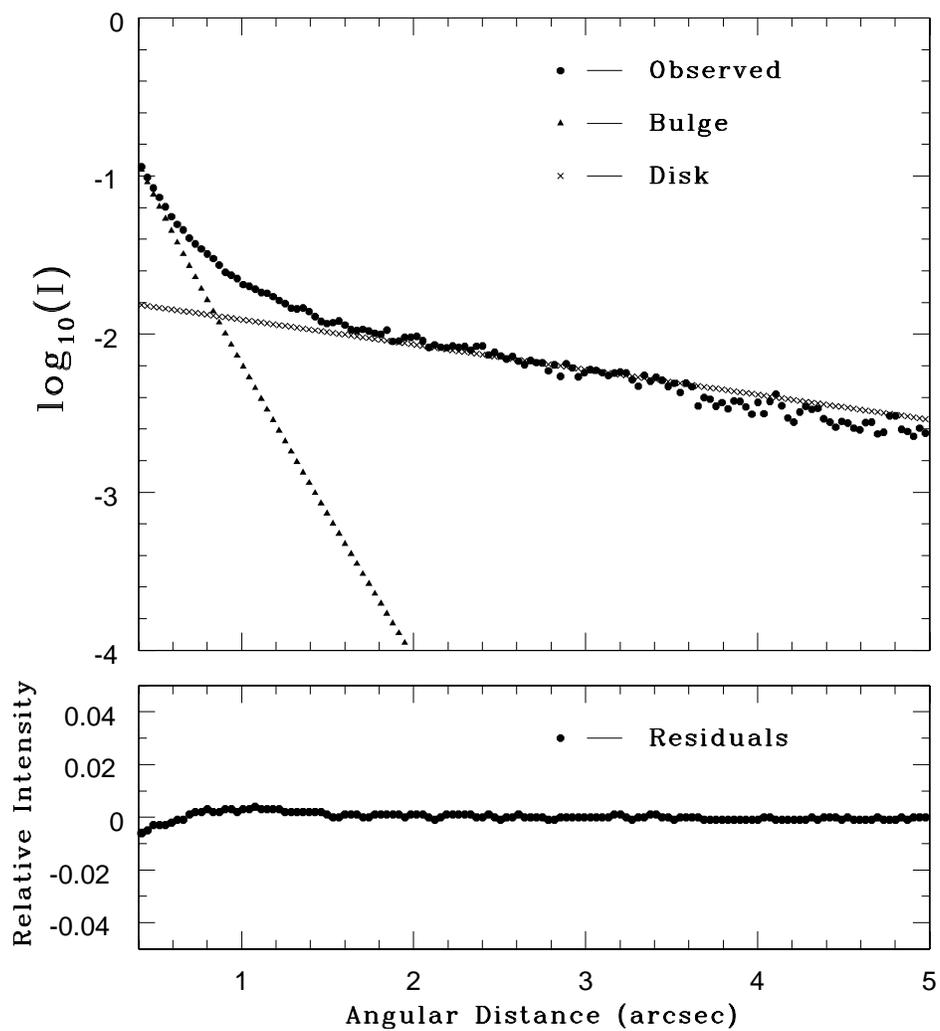} \figcaption{{\it J}-band surface brightness profile
for Mrk 618.}
\end{figure}
\begin{figure}
\plotone{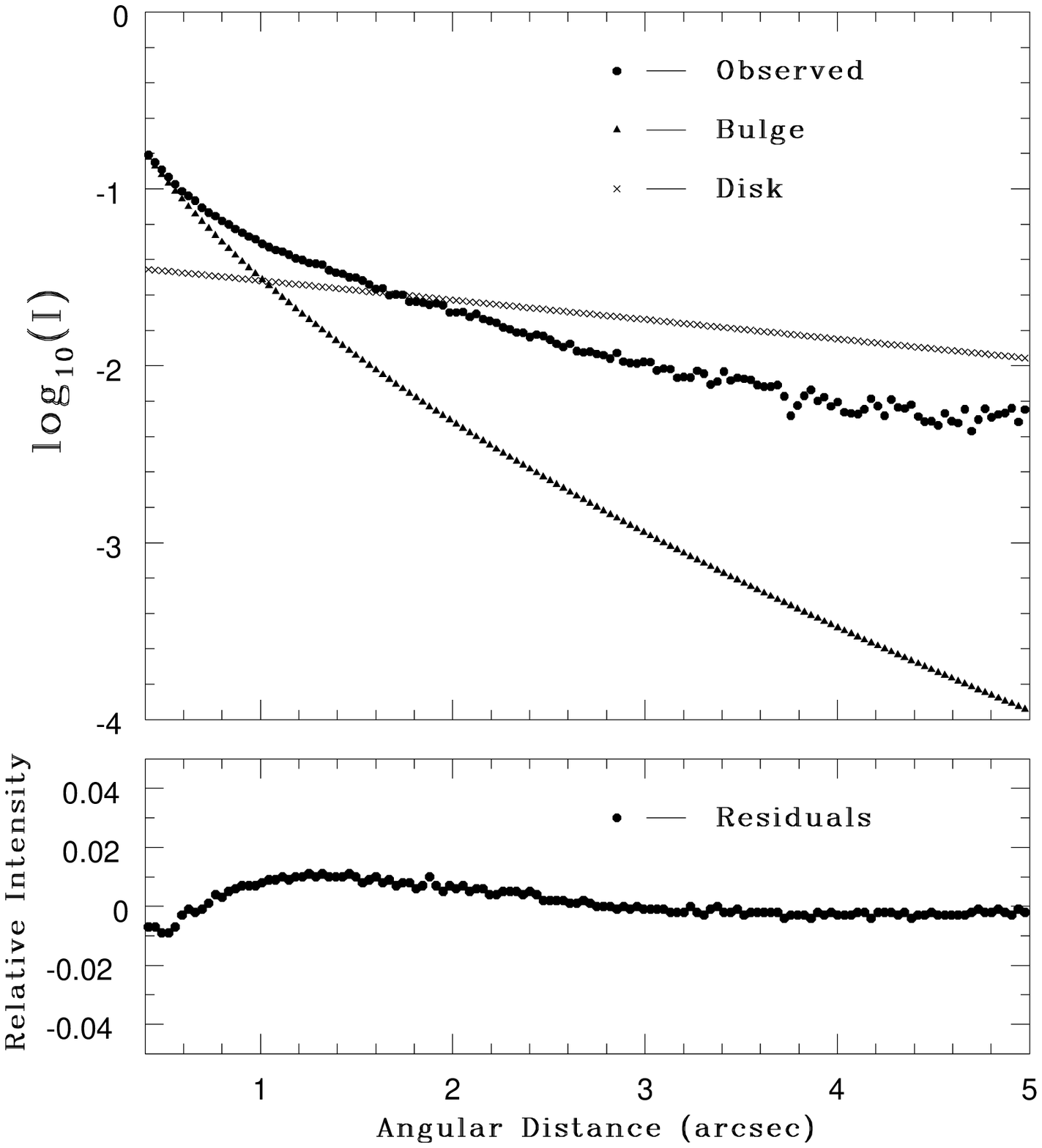} \figcaption{{\it J}-band surface brightness profile
for Mrk 359.}
\end{figure}
\begin{figure}
\plotone{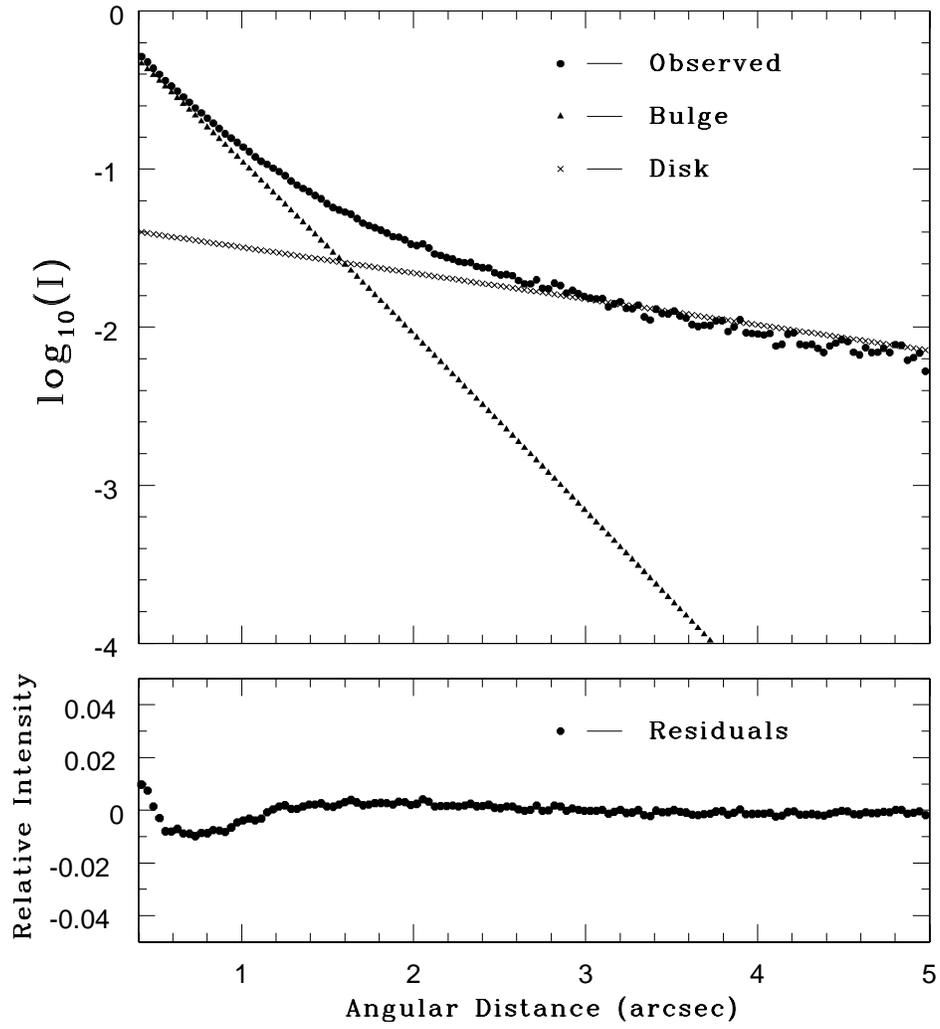} \figcaption{{\it J}-band surface brightness profile
for Mrk 1044.}
\end{figure}
\begin{figure}
\plotone{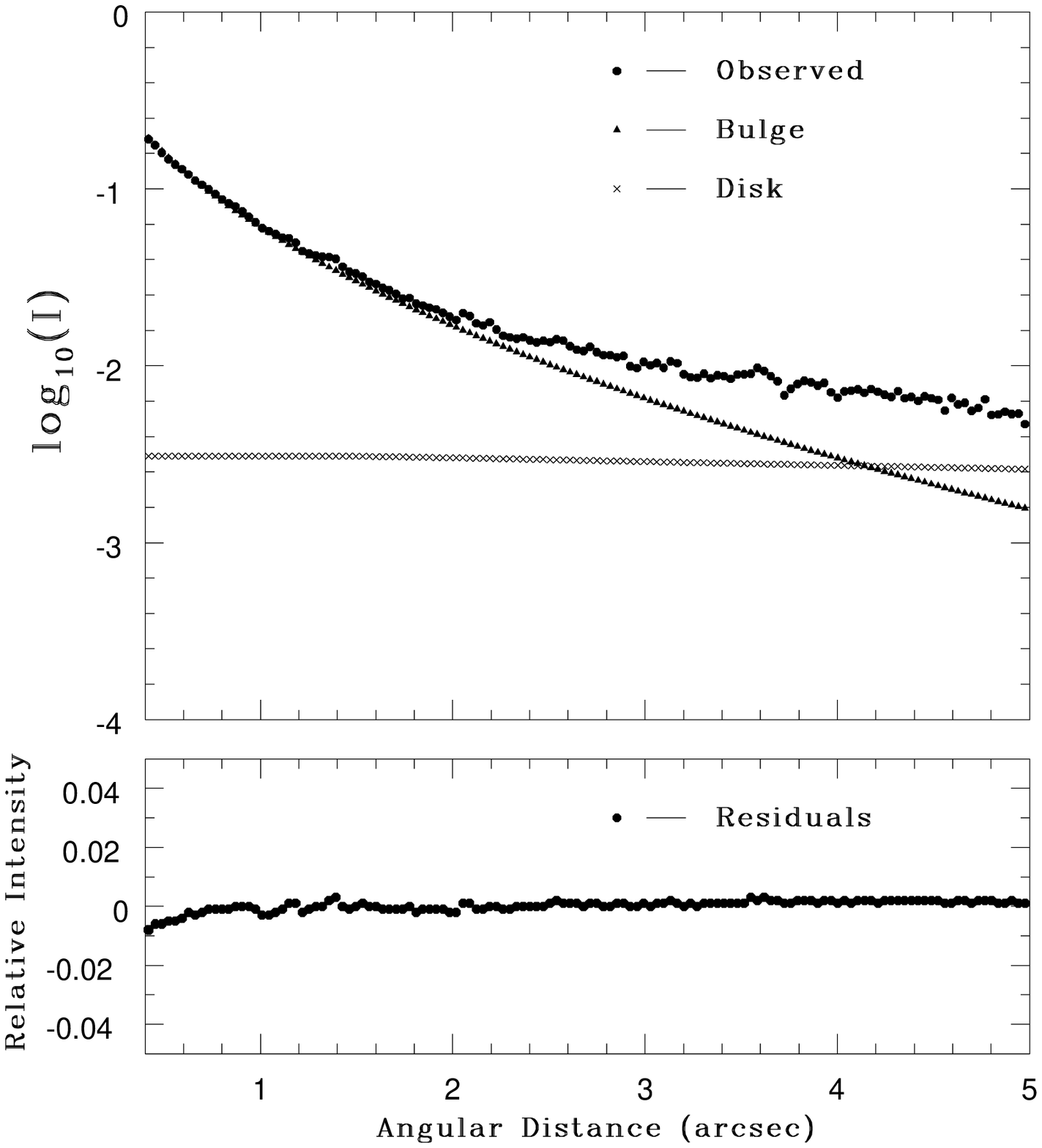} \figcaption{{\it J}-band surface brightness profile
for Mrk 705.}
\end{figure}
\begin{figure}
\plotone{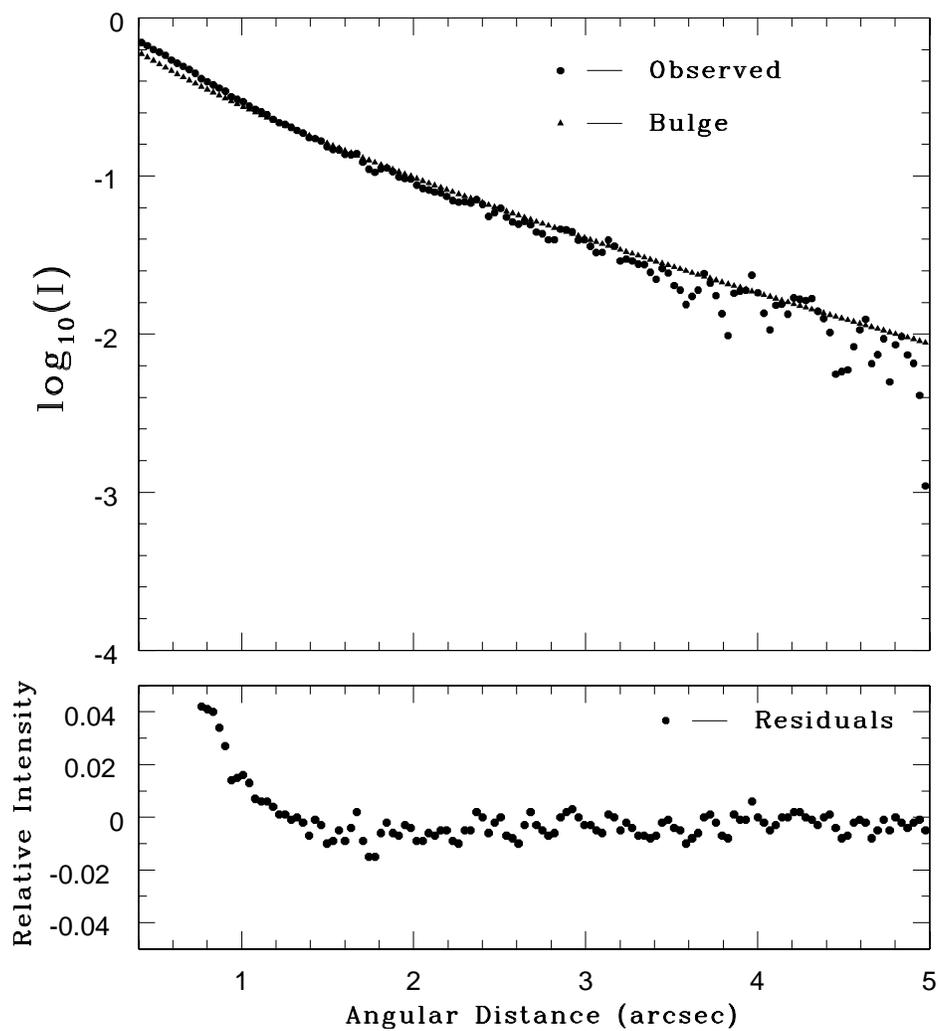} \figcaption{{\it J}-band surface brightness profile
for MCG 08.15.009.}
\end{figure}
\begin{figure}
\plotone{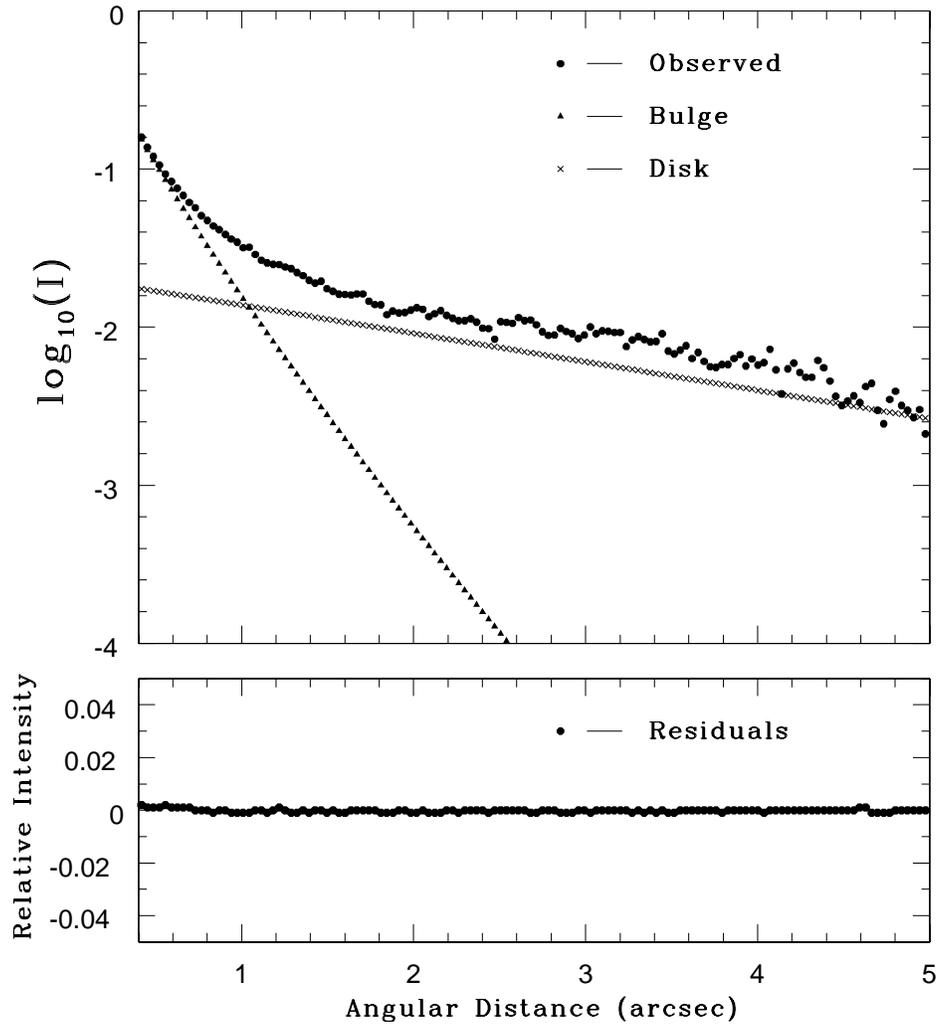} \figcaption{{\it J}-band surface brightness profile
for PG 1016+336.}
\end{figure}
\begin{figure}
\plotone{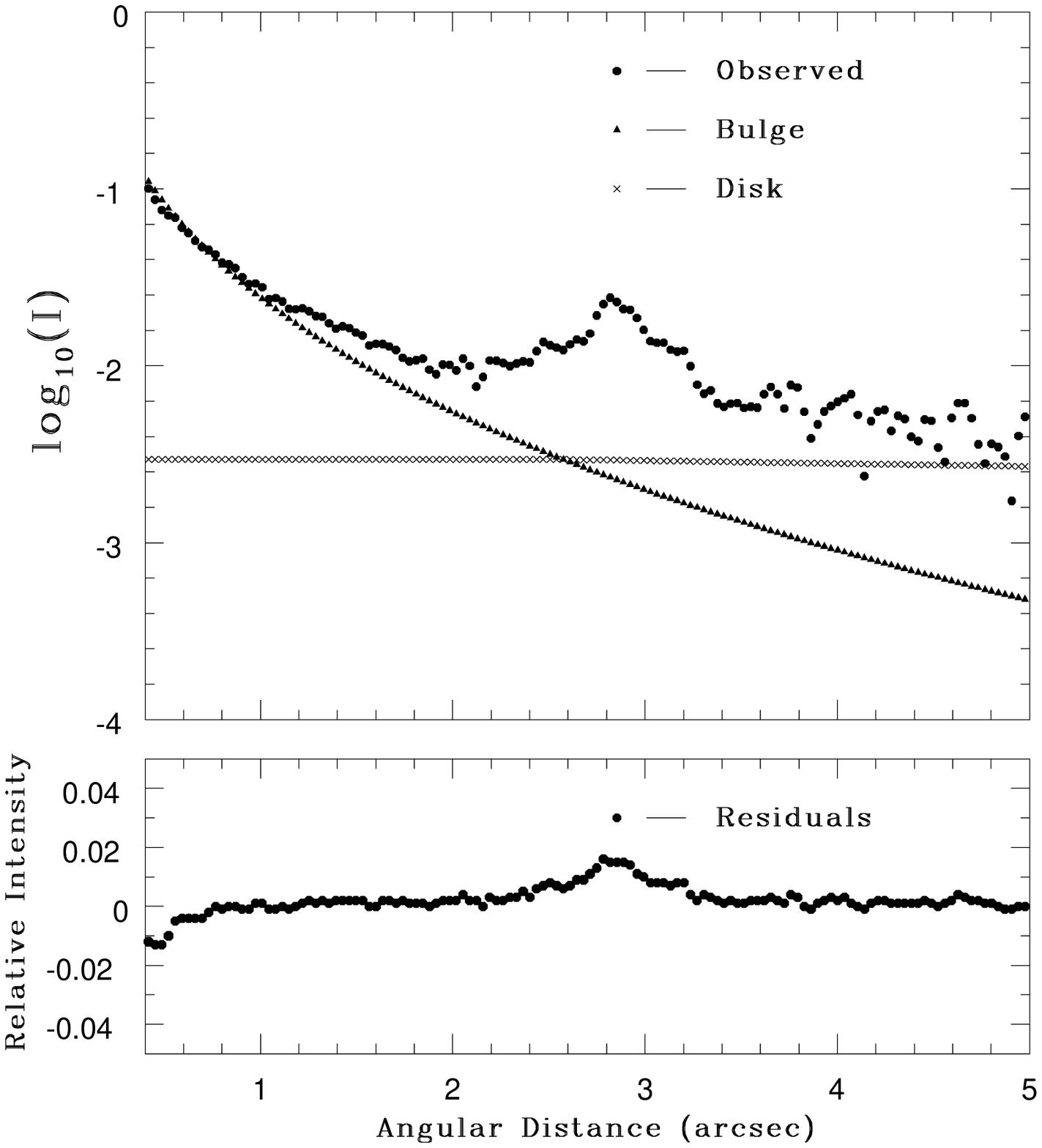} \figcaption{{\it J}-band surface brightness profile
for Mrk 734.}
\end{figure}


\clearpage

\begin{thebibliography}{}

\bibitem[Aguerri et al.(2005)]{agu05} Aguerri, J.A.L., Elias-Rosa, N.,
    Corsini, E.M., \& Mu{\~n}oz-Tu{\~n}{\'o}n, C. 2005, \aap, 434, 109
\bibitem[Andredakis, Peletier, \& Balcells(1995)]{and95} Andredakis,
    Y.C., Peletier, R.F., \& Balcells, M. 1995, \mnras, 275, 874
\bibitem[Balcells et al.(2003)]{bal03} Balcells, M., Graham, A.W., 
    Dom{\'i}nguez-Palmero, \& Peletier, R.F. 2003, \apjl, 582, L79
\bibitem[Barnes \& Hernquist(1991)]{bar91} Barnes, J.E. \& Hernquist, L.E.
    1991, \apjl, 370, L65
\bibitem[Barth, Greene, \& Ho(2005)]{bar05} Barth, A.J., Greene, J.E.,
    \& Ho, L.C. 2005, \apjl, 619, L151
\bibitem[Bender et al.(2005)]{ben05} Bender, R., et al. 2005, \apj, 631,
    280
\bibitem[Beuzit \& Hainaut(1998)]{beu98} Beuzit, J.-L. \& Hainaut,
    M.-C. 1998, User's Manual for the CFHT Adaptive Optics Bonnette, v0.9,
    The Canada-France-Hawaii Telescope Corporation 
\bibitem[Bian \& Zhao(2004)]{bia04} Bian, W. \& Zhao, Y. 2004,
    \mnras, 347, 607
\bibitem[Boller, Brandt, \& Fink(1996)]{bol96} Boller, Th., Brandt, W.N.,
    \& Fink, H. 1996, \aap, 305, 53
\bibitem[Boroson \& Green(1992)]{bor92} Boroson, T.A. \& Green,
    R.F. 1992, \apjs, 80, 109
\bibitem[Botte et al.(2004)]{bot04} Botte, V., Ciroi, S., Rafanelli,
    P., \& Di Mille, F. 2004, \aj, 127, 3168
\bibitem[Caon, Capaccioli, \& D'Onofrio(1993)]{cao93} Caon, N.,
    Capaccioli, M., \& D'Onofrio, M. 1993, \mnras, 265, 1013
\bibitem[Crenshaw, Kraemer \& Gabel(2003)]{cre03} Crenshaw, D.M., Kraemer,
    S.B., \& Gabel, J.R. 2003, \aj, 126, 1690
\bibitem[de Vaucouleurs(1948)]{dev48} de Vaucouleurs, G. 1948, Ann.
    Astrophys., 11, 247
\bibitem[de Vaucouleurs et al.(1991)]{dev91} de Vaucouleurs, G., de
    Vaucouleurs, A., Corwin, H.G., Buta, R.J., Paturel, G., \& Fouque,
    P. 1991, Third Reference Catalogue of Bright Galaxies (New York:
    Springer-Verlag)
\bibitem[Di Matteo, Springel, \& Hernquist(2005)]{dim05} Di Matteo, T.,
    Springel, V., \& Hernquist, L. 2005, \nat, 433, 604
\bibitem[Dong \& De Robertis(2006)]{don06} Dong, X.Y. \& De Robertis, M.M.
    2006, \aj, 131, 1236
\bibitem[Falc{\'o}n-Barroso, Peletier, \& Balcells(2002)]{fal02}
    Falc{\'o}n- Barroso, J., Peletier, R.F., \& Balcells, M. 2002,
    \mnras, 335, 741
\bibitem[Ferrarese \& Ford(2005)]{fer05} Ferrarese, L. \& Ford,
    H. 2005, \ssr, 116, 523
\bibitem[Ferrarese \& Merritt(2000)]{fer00} Ferrarese, L. \& Merritt,
    D. 2000, \apjl, 539, L9
\bibitem[Fioc \& Rocca-Volmerange(1999)]{fio99} Fioc, M. \& Rocca-Volmerange,
    B. 1999, \aap, 351, 869
\bibitem[Freeman(1970)]{fre70} Freeman, K.C. 1970, \apj, 160, 811
\bibitem[Gebhardt et al.(2000)]{geb00} Gebhardt, K., et al. 2000,
    \apjl, 539, L13
\bibitem[Graham(2001)]{gra01a} Graham, A.W. 2001, \aj, 121, 820
\bibitem[Greene \& Ho(2005a)]{gre05a} Greene, J.E. \& Ho, L.C. 2005a,
    \apj, 627, 721
\bibitem[Greene \& Ho(2005b)]{gre05b} Greene, J.E. \& Ho, L.C. 2005b,
    \apj, 630, 122
\bibitem[Grupe \& Mathur(2004)]{gru04a} Grupe, D. \& Mathur, S. 2004,
    \apjl, 606, L41
\bibitem[Grupe et al.(2004)]{gru04b} Grupe, D., Wills, B.J., Leighly, K.M.,
    \& Meusinger, H. 2004, \aj, 127, 156
\bibitem[Haehnelt \& Rees(1993)]{hae93} Haehnelt, M.G. \& Rees, M.J. 1993, 
    \mnras, 263, 168
\bibitem[H{\"a}ring \& Rix(2004)]{har04} H{\"a}ring, N. \& Rix,
    H.-W. 2004, \apjl, 604, L89
\bibitem[Hunt et al.(1997)]{hun97} Hunt, L.K., Malkan, M.A., Salvati, M.,
    Mandolesi, N., Palazzi, E., \& Wade, R. 1997, \apjs, 108, 229
\bibitem[Kaspi et al.(2005)]{kas05} Kaspi, S., Maoz, D., Netzer, H.,
Peterson, B.M., Vestergaard, M., \& Jannuzi, B.T. 2005, \apj, 629, 61
\bibitem[Kaspi et al.(2000)]{kas00} Kaspi, S., Smith, P.S., Netzer, H.,
    Maoz, D., Jannuzi, B.T., \& Giveon, U. 2000, \apj, 533, 631
\bibitem[Kim et al.(2005)]{kim05} Kim, S.S., Figer, D.F., Lee, M.G., \& Oh, S.
    2005, \pasp, 117, 445
\bibitem[Kormendy \& Illingworth(1982)]{kor82} Kormendy, J. \&
    Illingworth, G. 1982, \apj, 256, 460
\bibitem[Kormendy \& Kennicutt(2004)]{kor04} Kormendy, J. \&
    Kennicutt, R.C.  2004, \araa, 42, 603
\bibitem[Kormendy \& Richstone(1995)]{kor95} Kormendy, J. \&
    Richstone, D. 1995, \araa, 33, 581
\bibitem[Laor(2001)]{lao01} Laor, A. 2001, \apj, 553, 677
\bibitem[Liu \& Kennicutt(1995)]{liu95} Liu, C.T. \& Kennicutt, R.C. 1995,
    \apj, 450, 547
\bibitem[Magorrian et al.(1998)]{mag98} Magorrian, J., et al. 1998,
    \aj, 115, 2285
\bibitem[Marconi \& Hunt(2003)]{mar03} Marconi, A. \& Hunt, L.K. 2003,
    \apjl, 589, L21
\bibitem[M\'{a}rquez et al.(2000)]{mar00} M\'{a}rquez et al. 2000,
    \aap, 360, 431
\bibitem[Mathis(1990)]{mat90} Mathis, J.S. 1990, \araa, 28, 37
\bibitem[Mathur(2000)]{mat00} Mathur, S. 2000, \mnras, 314, L17
\bibitem[Mathur, Kuraszkiewicz, \& Czerny(2001)]{mat01} Mathur, S.,
    Kuraszkiewicz, J., \& Czerny, B. 2001, NewA, 6, 321
\bibitem[McLure \& Dunlop(2001)]{mcl01} McLure, R.J. \& Dunlop,
    J.S. 2001, \mnras, 327, 199
\bibitem[Nelson et al.(2004)]{nel04} Nelson, C.H., Green, R.F., Bower,
    G., Gebhardt, K., \& Weistrop, D. 2004, \apj, 615, 652
\bibitem[Osterbrock(1993)]{ost93} Osterbrock, D.E. 1993, \apj, 404,
    551
\bibitem[Osterbrock \& Pogge(1985)]{ost85} Osterbrock, D.E. \& Pogge,
    R.W.  1985, \apj, 297, 166
\bibitem[Peng et al.(2002)]{pen02} Peng, C.Y., Ho, L.C., Impey, C.D.,
    \& Rix, H.-W. 2002, \aj, 124, 266
\bibitem[Peterson \& Wandel(2000)]{pet00a} Peterson, B.M. \& Wandel,
    A. 2000, \apjl, 540, L13
\bibitem[Peterson et al.(2000)]{pet00b} Peterson, B.M., et al. 2000,
    \apj, 542, 161
\bibitem[Poggianti(1997)]{pog97} Poggianti, B.M. 1997, \aap,
    122, 399
\bibitem[Pounds, Done, \& Osborne(1995)]{pou95} Pounds, K.A., Done,
    C., \& Osborne, J. 1995, \mnras, 277, L5
\bibitem[Press et al.(1992)]{pre92} Press, W.H., Teukolsky, S.A.,
    Vetterling, W.T., \& Flannery, B.P. 1992, Numerical Recipes in C
    (2nd Ed.; Cambridge: Cambridge Univ. Press)
\bibitem[Rees(1984)]{ree84} Rees, M.J. 1984, \araa, 22, 471
\bibitem[Rigaut et al.(1998)]{rig98} Rigaut, F., et al. 1998, \pasp,
    110, 152
\bibitem[Rodr\'{i}guez-Ardila, Contini, \& Viegas(2005)]{rod05}
    Rodr\'{i}guez-Ardila, A., Contini, M., \& Viegas, S.M. 2005,
    \mnras, 357, 220
\bibitem[Rodr\'{i}guez-Ardila \& Mazzalay(2006)]{rod06}
    Rodr\'{i}guez-Ardila, A.  \& Mazzalay, X. 2006, \mnras, 367, L57
\bibitem[Rudy et al.(2000)]{rud00} Rudy, R.J., Mazuk, S., Puetter,
    R.C., \& Hamann, F. 2000, \apj, 539, 166
\bibitem[Schlegel, Finkbeiner, \& Davis(1998)]{sch98} Schlegel, D.J., 
    Finkbeiner, D.P., \& Davis, M. 1998, \apj, 500, 525
\bibitem[S\'{e}rsic(1968)]{ser68} S\'{e}rsic, J.L. 1968, Atlas de
    galaxias australes.  Observatorio Astronomico, Cordoba
\bibitem[Shemmer et al.(2001)]{she01} Shemmer, O., et al. 2001,
    \apj, 561, 162
\bibitem[Simien \& de Vaucouleurs(1986)]{sim86} Simien, F. \& de
    Vaucouleurs, G. 1986, \apj, 302, 564
\bibitem[Sulentic et al.(2000)]{sul00} Sulentic, J.W., Zwitter, T.,
    Marziani, P., \& Dultzin-Hacyan, D. 2000, \apjl, 536, L5
\bibitem[Tonry et al.(2001)]{ton01} Tonry, J.L., Dressler, A.,
    Blakeslee, J.P., Ajhar, E.A., Fletcher, A.B., Luppino, G.A.,
    Metzger, M.R., \& Moore, C.B.  2001, \apj, 546, 681
\bibitem[Vaduvescu \& McCall(2004)]{vad04} Vaduvescu, O. \& McCall,
    M.L. 2004, \pasp, 116, 640
\bibitem[Vanzi, Alonso-Herrero, \& Rieke(1998)]{van98} Vanzi, L.,
    Alonso-Herrero, A., \& Rieke, G.H. 1998, \apj, 504, 93
\bibitem[V{\'e}ron-Cetty \& V{\'e}ron(2001)]{ver01a} V{\'e}ron-Cetty,
    M.-P.  \& V{\'e}ron, P. 2001, \aap, 374, 92
\bibitem[V{\'e}ron-Cetty, V{\'e}ron, \& Gon\c{c}alves(2001)]{ver01b}
    V{\'e}ron-Cetty, M.-P., V{\'e}ron, P., \& Gon\c{c}alves,
    A.C. 2001, \aap, 372, 730
\bibitem[Virani, De Robertis, \& VanDalfsen(2000)]{vir00} Virani, S.N.,
    De Robertis, M.M., \& VanDalfsen, M.L. 2000, \aj, 120, 1739
\bibitem[Wainscoat \& Cowie(1992)]{wai92} Wainscoat, R.J. \& Cowie,
    L.L.  1992, \aj, 103, 332
\bibitem[Wandel(2002)]{wan02} Wandel, A. 2002, \apj, 565, 762
\bibitem[Wang \& Lu(2001)]{wan01} Wang, T. \& Lu, Y. 2001, \aap 377, 52
\bibitem[Williams, Pogge, \& Mathur(2002)]{wil02} Williams, R.J.,
    Pogge, R.W., \& Mathur, S. 2002, \aj, 124, 3042
\bibitem[York et al.(2000)]{yor00} York, D.G. et al. 2000, \aj, 120,
    1579
\end{thebibliography}
\end{document}